\documentclass[12pt]{article}

\usepackage[margin=1in]{geometry}
\usepackage{amsmath,amssymb,amsfonts,amsthm}
\theoremstyle{definition}
\newtheorem{remark}{Remark}
\usepackage{booktabs}
\usepackage{graphicx}
\usepackage{natbib}
\usepackage{algorithm}
\usepackage{algpseudocode}
\usepackage[colorlinks=True,citecolor={blue}]{hyperref}
\usepackage{multirow}
\usepackage{xcolor}
\usepackage[normalem]{ulem}

\makeatletter
\g@addto@macro\@floatboxreset{\normalcolor}
\g@addto@macro\fst@algorithm{%
  \edef\@fs@pre{\noexpand\normalcolor\unexpanded\expandafter{\@fs@pre}}%
  \edef\@fs@mid{\noexpand\normalcolor\unexpanded\expandafter{\@fs@mid}}%
  \edef\@fs@post{\noexpand\normalcolor\unexpanded\expandafter{\@fs@post}}}
\makeatother

\providecommand{\markuprevision}{0}
\definecolor{revadded}{RGB}{0,70,200}
\definecolor{revdeleted}{RGB}{200,0,0}
\makeatletter
\newcommand{\revtag}[1]{\ifx\relax#1\relax\else\ifmmode\else\ifx\@captype\@undefined\ifinner\else\marginpar{\raggedright\tiny\textcolor{revadded}{#1}}\fi\fi\fi\fi}
\makeatother
\if1\markuprevision
  \hypersetup{colorlinks=false,pdfborder={0 0 0}}
  \newcommand{\added}[2][]{\revtag{#1}{\color{revadded}#2}}
  \newcommand{\deleted}[2][]{\ifmmode\else\revtag{#1}{\color{revdeleted}\sout{#2}}\fi}
  \newcommand{\deletedblock}[2][]{\revtag{#1}{\color{revdeleted}#2}}
  \newcommand{\replaced}[3][]{\ifmmode{\color{revadded}#2}\else\revtag{#1}{\color{revadded}#2}\ {\color{revdeleted}\sout{#3}}\fi}
\else
  \newcommand{\added}[2][]{#2}
  \newcommand{\deleted}[2][]{}
  \newcommand{\deletedblock}[2][]{}
  \newcommand{\replaced}[3][]{#2}
\fi

\title{Robust Twoblock Simultaneous Dimension Reduction\added{ and Multivariate Regression}\footnote{Please cite this article as: S. Serneels, {\em Journal of Chemometrics}, 40 (2026), e70195, \href{http://doi.org/10.1002/cem.70195}{http://doi.org/10.1002/cem.70195}.}}
\author{Sven Serneels$^{1,2}$ \\[6pt]
\small $^1$ Snow Stallion AI, Cheyenne, Wyoming, USA \\
\small $^2$ Department of Mathematics, University of Antwerp, Belgium\footnote{Corresponding author, e-mail: \href{mailto:svenserneels@gmail.com}{svenserneels@gmail.com}.}}
\date{}

\begin{document}

\maketitle

\if1\markuprevision
\begin{center}
\fbox{\parbox{0.9\textwidth}{\small Revised manuscript with changes marked:
\textcolor{revadded}{added text in blue}, \textcolor{revdeleted}{\sout{deleted text in red}}.
Margin labels refer to the numbered reviewer comments in the response document.}}
\end{center}
\fi

\begin{abstract}
This paper introduces robust twoblock (RTB) simultaneous dimension reduction, which is the first statistically robust method to perform simultaneous dimension reduction in two blocks of variables \replaced[R1.1, R1.2]{that allows to tune the number of components in each block individually.}{and allows to fine-tune the model complexity in each block individually.} The paper proposes both a dense and a sparse version of the new method. Sparse RTB is the first robust estimator \replaced[R1.2]{of this kind that allows to tune both the number of components and the degree of sparsity (i.e.\ the number of variables retained) for each block individually.}{that allows to select both model complexity and the degree of sparsity for each block individually.} RTB thereby allows to optimally extract and summarize the relevant portion of information in each block of data, also in the presence of outliers. As a corollary, the estimators can be recombined into a single estimate of regression coefficients for multivariate regression that is operable when the number of variables exceeds the number of cases in each block. An extensive simulation study illustrates that the new methods are resistant to different types of outliers, while maintaining estimation efficiency across a range of dimensionality settings. These findings both hold true for the dense and the sparse method. The methods' performance is further illustrated on two example data sets and a straightforward algorithm is presented and made accessible in an open source repository.  
\end{abstract}


\section{Introduction}
\added[R1.D]{\subsection{Background and motivation}\label{sec:background}}

As data dimensionality continues to grow, the ability to compress information into a compact low-dimensional representation becomes ever more critical. When data are structured as two co-observed blocks of variables, this compression should respect the inter-block relationship: the reduced spaces ought to capture all information relevant to model each block from the other. Simultaneous two-block sufficient dimension reduction (SDR) provides exactly this, yielding a principled and parsimonious
summary of the joint structure of both blocks.

Formally, let $\mathbf{X} \in \mathbb{R}^{n \times p}$ and $\mathbf{Y} \in \mathbb{R}^{n \times q}$ denote matrices of $n$ observations of $p$-variate independent and $q$-variate dependent random variables, respectively. SDR seeks \replaced[R1.4]{components (or scores), i.e.\ linear combinations}{latent variables} $\mathbf{T} = \mathbf{X}\mathbf{W}$\added[R1.4]{ of the original variables,} satisfying
\begin{equation}\label{eq:1bsdr}
    \mathbf{Y} \perp\!\!\!\perp \mathbf{X} \mid \mathbf{T},
\end{equation}
where the fulfilling subspace is called the \emph{central subspace} and $\perp\!\!\!\perp$ denotes statistical independence. \added[R1.4]{In the PLS literature, such components are commonly referred to as latent variables. Strictly speaking, however, they are not latent variables as commonly understood in the path modelling and multiblock literature, where a latent variable is an unobserved quantity that generates the observed variables (a reflective measurement model). Being linear combinations of the observed variables, the components are composites instead. In this paper, they are therefore referred to as components or scores.} Among estimators of the central subspace, partial least squares (PLS) remains one of the oldest and most computationally efficient, with its theoretical grounding well established. More specifically, \cite{cook2013} derived that PLS1 (univariate response) provides a consistent estimator of the central subspace $\mathcal{S}_{\mathbf{Y}|\mathbf{X}}$ when $\mathbf{Y}$ is univariate. It accomplishes this because the PLS1 weight vectors span a Krylov subspace of $\boldsymbol{\Sigma}_{\mathbf{XX}}^{-1} \boldsymbol{\Sigma}_{\mathbf{XY}}$, which under mild conditions converges to $\mathcal{S}_{\mathbf{Y}|\mathbf{X}}$.

Equation \eqref{eq:1bsdr} addresses dimension reduction in $\mathbf{X}$ alone. A richer objective is to simultaneously reduce both blocks, seeking linear combinations $\mathbf{U} = \mathbf{Y}\mathbf{V}$
and $\mathbf{T} = \mathbf{X}\mathbf{W}$ such that
\begin{subequations}\label{eq:sdr2}
\begin{align}
    \mathbf{Y} &\perp\!\!\!\perp \mathbf{X} \mid \mathbf{T}, \label{eq:sdr_Y}\\
    \mathbf{X} &\perp\!\!\!\perp \mathbf{Y} \mid \mathbf{U}. \label{eq:sdr_X}
\end{align}
\end{subequations}
\added[R2.1, R2.2]{Condition~\eqref{eq:sdr_Y} states that $\mathbf{T}$ retains all information in $\mathbf{X}$ that is relevant to $\mathbf{Y}$, whereas condition~\eqref{eq:sdr_X} states that $\mathbf{U}$ retains all information in $\mathbf{Y}$ that is relevant to $\mathbf{X}$. Both conditions are reductions; neither implies that one block is to be predicted from the other. In this paper, as in \cite{cook2023}, the simultaneous reduction is used to estimate a regression of $\mathbf{Y}$ on $\mathbf{X}$: the reduced $\mathbf{Y}$-subspace enters the coefficient estimate through the $\mathbf{Y}$ weights $\mathbf{V}$ (Algorithm~\ref{alg:twoblock}), but predicting $\mathbf{X}$ from $\mathbf{Y}$ is not pursued.} 
Since independence is difficult to establish directly, these conditions are relaxed to zero covariance. Letting
$\mathbf{P} = \mathbf{X}^T\mathbf{T}(\mathbf{T}^T\mathbf{T})^{-1}$
and
$\mathbf{Q} = \mathbf{Y}^T\mathbf{U}(\mathbf{U}^T\mathbf{U})^{-1}$
denote the $\mathbf{X}$- and $\mathbf{Y}$-loadings, with residual complements
$\mathbf{E} = \bigl(\mathbf{I}_n -
\mathbf{T}(\mathbf{T}^T\mathbf{T})^{-1}\mathbf{T}^T\bigr)\mathbf{X}$
and
$\mathbf{F} = \bigl(\mathbf{I}_n -
\mathbf{U}(\mathbf{U}^T\mathbf{U})^{-1}\mathbf{U}^T\bigr)\mathbf{Y}$,
\added[R1.5]{the pairs $(\mathbf{P}, \mathbf{E})$ and $(\mathbf{Q}, \mathbf{F})$ each pertain to a single block: $\mathbf{P}$ contains the coefficients of the least squares regression of $\mathbf{X}$ on its own scores $\mathbf{T}$, such that $\mathbf{X} = \mathbf{T}\mathbf{P}^T + \mathbf{E}$, with $\mathbf{E}$ the part of $\mathbf{X}$ that is not captured by the reduced $\mathbf{X}$ space. Likewise, $\mathbf{Q}$ contains the coefficients of the regression of $\mathbf{Y}$ on its own scores $\mathbf{U}$, such that $\mathbf{Y} = \mathbf{U}\mathbf{Q}^T + \mathbf{F}$, with $\mathbf{F}$ the part of $\mathbf{Y}$ that is not captured by the reduced $\mathbf{Y}$ space. Note that $\mathbf{F}$ is therefore not the residual of a regression of $\mathbf{Y}$ on $\mathbf{X}$. With this notation,} simultaneous dimension reduction of both blocks requires
\begin{subequations}\label{eq:SDRcov}
\begin{align}
    \mathrm{cov}(\mathbf{E},\, \mathbf{F}) &= \mathbf{0}, \label{eq:cov_EF}\\
    \mathrm{cov}(\mathbf{U}\mathbf{Q}^T,\, \mathbf{F}) &= \mathbf{0}, \label{eq:cov_UF}\\
    \mathrm{cov}(\mathbf{T}\mathbf{P}^T,\, \mathbf{E}) &= \mathbf{0}. \label{eq:cov_TE}
\end{align}
\end{subequations}

Partial least squares (PLS) regression is also a widely used method for simultaneous dimension reduction in a multivariate regression context. It is often silently assumed that multivariate PLS (also denoted PLS2) yields an estimate for both central subspaces in \eqref{eq:SDRcov}.
Howbeit, while \cite{cook2013} proved that the PLS1 score space is an estimator for the central subspace $\mathcal{S}_{\mathbf{Y}|\mathbf{X}}$, \cite{cook2023} showed that this result does \emph{not} extend to PLS2 (multivariate response). PLS2 only reduces the $\mathbf{X}$ block, while treating $\mathbf{Y}$ as fixed, and consequently does not estimate the joint two-block central subspace~\eqref{eq:sdr2}.
This motivated their development of \emph{two-block simultaneous dimension reduction} (hereafter: twoblock, also referred to as ``XY-PLS"), which simultaneously extracts \replaced[R1.4]{components}{latent components} from both the predictor block $\mathbf{X}$ and the response block $\mathbf{Y}$, providing an estimator for the two-block SDR problem. The method proceeds in two sequential deflation stages: first extracting $h_x$ components from $\mathbf{X}$ that are maximally covariant with $\mathbf{Y}$ and then extracting $h_y$ components from $\mathbf{Y}$ that are maximally covariant with $\mathbf{X}$. This provides true simultaneous SDR estimates for both blocks, along with direct regression coefficient estimates. It was shown to outperform both PLS2 and various variants of univariate regression\added[R2.3]{: in simulations, in terms of both the mean squared prediction error and the mean squared estimation error of the regression coefficients, and on three real data sets in terms of the mean squared prediction error \citep{cook2023}. Compared methods included ordinary least squares, PLS2, PLS1, Wold's two-block algorithm and maximum likelihood envelope estimation, and twoblock was judged to perform best overall}. 

The author \citep{serneels2025} has recently introduced sparse twoblock dimension reduction, which is a sparse extension of the original method, by \replaced{eliminating variables with small weights from the weight vectors}{applying soft-thresholding to the weight vectors} during each deflation step.
This produces variable selection in both the $\mathbf{X}$ and $\mathbf{Y}$ blocks, yielding interpretable models with competitive predictive performance. \replaced[R1.1, R1.7]{Unlike sparse PLS2 \citep{chun2010} and sparse canonical correlation analysis (CCA), which use a single number of components for both blocks, it allows both the number of components and the sparsity to be tuned in each block of variables individually, within a simultaneous dimension reduction framework that yields a regression model for $\mathbf{Y}$ given $\mathbf{X}$. Its relation to other multiblock methods is discussed in Section~\ref{sec:related}.}{Beyond being a versatile alternative to both sparse PLS2 and sparse canonical correlation analysis (CCA), it is also the first method to allow fine-tuning of both the model complexity and the sparsity in each block of variables individually.}

Both the dense and sparse twoblock methods rely on classical location and shape estimates, making them sensitive to outlying observations. In chemometrics, along other applied domains, data frequently contain atypical cases, such as leverage points in the predictor space, vertical outliers in the response, or both. The presence of such atypical cases that can severely distort the estimated subspaces and regression coefficients. With that in mind, this paper introduces \emph{robust twoblock} (RTB) dimension reduction. The RTB algorithm proposed herein applies iterative reweighting, inherited from M-estimation. For a detailed overview of robust estimates in a regression context, we refer to e.g. \cite{maronna2019robust}. This approach to robustifying dimension reduction and regression algorithms was previously successfully implemented by the author in the context of partial least squares, leading to the now widely adopted methods of Partial Robust M-regression (PRM, \cite{serneels2005}) and its sparse extension SPRM \citep{hoffmann2015}. The algorithm introduced herein builds upon those two robust estimators, but introduces multivariate reweighting scheme adapted to the twoblock setting. Caseweights are computed independently for the $\mathbf{X}$ and $\mathbf{Y}$ blocks based on Mahalanobis-type distances in the respective score spaces, then used to downweight or omit atypical cases when estimating the twoblock decomposition. The resulting method is the first robust version of twoblock simultaneous dimension reduction altogether, a statement that applies both to the dense and sparse variants.

The remainder of this paper is organized as follows. \added[R1.D]{Section~\ref{sec:related} positions RTB with respect to related two-block and multiblock methods.} Section~\ref{sec:method} introduces the new RTB method, along with an algorithm that can be implemented straightforwardly. Section~\ref{sec:simulation} presents a simulation study evaluating RTB under various contamination and dimensionality scenarios. Section~\ref{sec:examples} illustrates the method on two real data sets that are known to benefit from multivariate estimation and contain outliers. Finally, Section~\ref{sec:conclusion} concludes and provides a perspective into pathways for further future research.

\added[R1.1, R1.3, R1.6--R1.8, R1.D]{%
\subsection{Related work}\label{sec:related}

Twoblock and RTB belong to a broad family of component-based methods that relate two or more blocks of variables measured on the same cases. Since several of these methods share elements with the proposal, this subsection outlines how RTB differs from them in four respects: (i) whether a reduced subspace is estimated for the response block $\mathbf{Y}$ as well as for the predictor block $\mathbf{X}$; (ii) whether the number of components can be set separately for each block; (iii) whether variable selection is possible with a degree of sparsity that can be tuned separately for each block; and (iv) whether the method is statistically robust, i.e.\ resistant to outlying cases. Table~\ref{tab:related} summarizes the comparison.

A first group of related methods consists of supervised methods that reduce the predictor block only. PLS2, reduced rank regression (RRR) and principal covariates regression (PCovR, \citealp{dejong1992}) extract components from $\mathbf{X}$ that summarize $\mathbf{X}$ and predict $\mathbf{Y}$. Continuum redundancy-PLS regression \citep{bougeard2008} connects redundancy analysis (i.e.\ RRR) and PLS2 through a ridge-type parameter that addresses multicollinearity, and multiblock redundancy analysis \citep{bougeard2011} extends this continuum to several explanatory blocks. In these methods, only the explanatory block(s) are deflated and a single number of components applies; $\mathbf{Y}$ is not itself reduced, and neither sparsity nor robustness against outliers is provided. Sparse PLS \citep{chun2010} adds variable selection in $\mathbf{X}$, again with a single set of components. Sparse multivariate principal covariates regression (SMPCovR, \citealp{park2024}) adds elastic net penalties both to the PCovR weights and to the loadings that regress $\mathbf{Y}$ on the components, with separate tuning parameters for each. It thereby selects predictors as well as outcomes: an outcome variable whose loadings are all shrunk to zero is excluded from the model. Among existing methods, SMPCovR is therefore the closest in spirit to sparse twoblock. However, SMPCovR builds a single set of components $\mathbf{T} = \mathbf{X}\mathbf{W}$ that is shared by both blocks, such that no separate reduced subspace for $\mathbf{Y}$ is estimated and the number of components cannot differ between both blocks. Moreover, like all methods in this group, it is based on a least squares criterion and is therefore not robust to outliers.

A second group consists of symmetric multiblock component methods, which treat all blocks on an equal footing. Regularized generalized canonical correlation analysis (RGCCA, \citealp{tenenhaus2011,tenenhaus2017}) is a unifying framework for component methods on two or more blocks. Through block-specific shrinkage parameters, a design matrix that specifies which blocks are connected and a scheme function, it includes PLS path modelling, Tucker's inter-battery factor analysis \citep{tucker1958}, redundancy analysis and CCA as special cases. Its sparse variant SGCCA \citep{tenenhaus2014} imposes an $\ell_1$ constraint with a separate sparsity parameter for each block. Higher order components are obtained by deflation, and \cite{tenenhaus2017} note that the number of components summarizing a block can vary from one block to another. RGCCA and SGCCA thus allow complexity and sparsity to be tuned per block. They are, however, exploratory methods that maximize a symmetric criterion of between-block association: they do not deliver a regression model for $\mathbf{Y}$ given $\mathbf{X}$, components are extracted jointly, stage by stage, for all blocks (a different number of components per block is obtained by excluding blocks from later stages, rather than from separate deflation sequences for each block as in twoblock), and they are not robust to outliers. The precise relation between the first twoblock components and RGCCA is stated in Remark~\ref{rem:rgcca}. Multiple co-inertia analysis \citep{chessel1996} and its recent simultaneous counterpart \citep{galharret2026} summarize several blocks by a common set of scores with block-specific loadings. In the latter, the attribute \emph{robust} refers to uniqueness and stability of the solution, which is a different notion from the statistical robustness against outliers that is pursued in this paper.

A third, related line of research decomposes multiblock data into components that are common to all blocks and components that are distinctive for one or a subset of blocks; see \cite{smilde2017} for an overview and \cite{park2021} for a sparse covariates regression method with several predictor blocks and a univariate outcome. Twoblock and RTB are not designed to separate common from distinctive variation. The simultaneous dimension reduction conditions~\eqref{eq:SDRcov} target the part of each block that is relevant to the other block, i.e.\ the redundancy between both blocks, whereas variation that is specific to either block is relegated to the residuals $\mathbf{E}$ and $\mathbf{F}$.

Finally, as far as robustness is concerned, robust counterparts exist for univariate PLS (PRM, \citealp{serneels2005}; SPRM, \citealp{hoffmann2015}) and for sparse CCA \citep{wilms2016}. Robust sparse CCA allows a separate sparsity parameter per block, but constrains both blocks to the same number of canonical variates. To the best of the author's knowledge, none of the methods discussed above has a statistically robust counterpart that simultaneously estimates reduced subspaces for both blocks, with a number of components and a degree of sparsity that can be tuned separately for each block. RTB is the first method to offer this combination.

\begin{table}[tbp]
\centering
\caption{Properties of RTB and related methods: (i) a reduced subspace is estimated for $\mathbf{Y}$ as well as for $\mathbf{X}$; (ii) the number of components can differ between blocks; (iii) the degree of sparsity can be tuned separately per block; (iv) the method yields a regression model for $\mathbf{Y}$ given $\mathbf{X}$; (v) the method is robust to outliers. References for each method are given in Section~\ref{sec:related}.}
\label{tab:related}
\small
\begin{tabular}{p{0.36\textwidth}ccccc}
\toprule
Method & (i) & (ii) & (iii) & (iv) & (v) \\
\midrule
PLS2, RRR, PCovR                 & -- & -- & -- & \checkmark & -- \\
Sparse PLS                       & -- & -- & $\mathbf{X}$ only & \checkmark & -- \\
Continuum RA--PLS, mbRA          & -- & -- & -- & \checkmark & -- \\
SMPCovR                          & --$^a$ & -- & \checkmark & \checkmark & -- \\
RGCCA / SGCCA                    & \checkmark & \checkmark$^b$ & SGCCA & -- & -- \\
MCOA / OMCOA                     & \checkmark$^c$ & -- & -- & -- & --$^d$ \\
Robust sparse CCA                & \checkmark & -- & \checkmark & -- & \checkmark \\
Twoblock / sparse twoblock       & \checkmark & \checkmark & sparse & \checkmark & -- \\
RTB / sparse RTB (this paper)    & \checkmark & \checkmark & sparse & \checkmark & \checkmark \\
\bottomrule
\end{tabular}

\smallskip
\parbox{0.8\textwidth}{\footnotesize $^a$ Outcome variables can be selected, but a single set of components is shared by both blocks. $^b$ By excluding blocks from later stages of the joint deflation. $^c$ A common set of scores with block-specific loadings. $^d$ \emph{Robust} in the sense of uniqueness and stability of the solution, not resistance to outliers.}
\end{table}
}

\section{Method}\label{sec:method}

Prior to introducing the new robust estimators, a brief overview will be provided of the existing non-robust alternatives. 
\subsection{Dense and sparse twoblock dimension reduction}

Let the predictor matrix $\mathbf{X}$ and predictand matrix $\mathbf{Y}$ both be centred and scaled.
The twoblock method \citep{cook2023} then extracts components sequentially. \replaced[R1.9]{It first extracts $h_x$ components from $\mathbf{X}$, each of which has maximal covariance with the original (undeflated) $\mathbf{Y}$ block, deflating $\mathbf{X}$ after each component. It then extracts $h_y$ components from $\mathbf{Y}$ in the same way, with the roles of both blocks switched. Formally, the $i$-th $\mathbf{X}$-block weight vector solves
\begin{equation}\label{eq:tbcrit}
    \mathbf{w}_i = \operatorname*{argmax}_{\|\mathbf{w}\| = 1} \bigl\|\mathbf{Y}_0^\top \tilde{\mathbf{X}}^{(i)} \mathbf{w}\bigr\|^2 = \operatorname*{argmax}_{\|\mathbf{w}\| = 1} \sum_{l=1}^{q} \mathrm{cov}^2\bigl(\tilde{\mathbf{X}}^{(i)}\mathbf{w}, \mathbf{y}_{0,l}\bigr),
\end{equation}
where $\tilde{\mathbf{X}}^{(i)}$ is the $\mathbf{X}$ block after deflation of the first $i-1$ components and $\mathbf{y}_{0,l}$ is the $l$-th column of $\mathbf{Y}_0$. This is the PLS2 criterion applied to the deflated $\mathbf{X}$ and the undeflated $\mathbf{Y}$ block, and its solution is the leading left singular vector of $\tilde{\mathbf{X}}^{(i)\top}\mathbf{Y}_0$. The $\mathbf{Y}$-block weight vectors $\mathbf{v}_j$ solve the analogous problem with $\mathbf{X}$ and $\mathbf{Y}$ switched. Since each weight vector follows directly from a singular value decomposition, no inner iterations are required; the algorithm shares with NIPALS \citep{wold1966} only its sequential deflation structure. \cite{cook2023} show that the conditions for simultaneous reduction of both blocks are satisfied by the predictor and response envelopes, i.e.\ the smallest reducing subspaces of $\boldsymbol{\Sigma}_{\mathbf{X}}$ and $\boldsymbol{\Sigma}_{\mathbf{Y}}$ that contain the column and row space of the regression coefficient matrix, respectively (their Proposition~5), and that the weight matrices obtained by the above procedure are $\sqrt{n}$-consistent estimators of bases of these envelopes (their Propositions~2 and~4). This makes twoblock an estimator for the two-block SDR problem.}{In essence, it applies a NIPALS \mbox{\citep{wold1966}} inspired algorithm to calculate latent variables to the original predictand and predictor data to extract the latent space for $\mathbf{X}$ and then applies the algorithm again to data in which the roles of predictor and predictand have been switched.} Eventually, both sets of estimates are recombined into a single set of regression coefficients. 

The algorithm is summarized in Algorithm~\ref{alg:twoblock}, where $\mathbf{W} = [\mathbf{w}_1, \ldots, \mathbf{w}_{h_x}]$ and $\mathbf{V} = [\mathbf{v}_1, \ldots, \mathbf{v}_{h_y}]$ collect the weight vectors for the $\mathbf{X}$ and $\mathbf{Y}$ blocks, respectively.

\begin{algorithm}[t]
\caption{Twoblock dimension reduction (optionally sparse)}\label{alg:twoblock}
\begin{algorithmic}[1]
\Require $\mathbf{X} \in \mathbb{R}^{n \times p}$, $\mathbf{Y} \in \mathbb{R}^{n \times q}$, components $h_x$, $h_y$
\State \replaced{$\mathbf{m}_X, \mathbf{m}_Y \gets$ column means of $\mathbf{X}$, $\mathbf{Y}$; \ $\mathbf{S}_X, \mathbf{S}_Y \gets$ diagonal matrices of column standard deviations ($\mathbf{I}$ if unscaled); \ $\mathbf{X}_0 \gets (\mathbf{X} - \mathbf{1}_n \mathbf{m}_X^\top)\mathbf{S}_X^{-1}$, \ $\mathbf{Y}_0 \gets (\mathbf{Y} - \mathbf{1}_n \mathbf{m}_Y^\top)\mathbf{S}_Y^{-1}$}{Centre and scale $\mathbf{X} \to \mathbf{X}_0$, $\mathbf{Y} \to \mathbf{Y}_0$}
\State $\tilde{\mathbf{X}}^{(1)} \gets \mathbf{X}_0$; \quad $\tilde{\mathbf{Y}}^{(1)} \gets \mathbf{Y}_0$\added{; \quad $\mathcal{A}_x, \mathcal{A}_y \gets \emptyset$}
\Statex \textit{--- $\mathbf{X}$-block deflation ---}
\For{$i = 1, \ldots, h_x$}
    \State $\mathbf{S}_{XY}^{(i)} \gets \tilde{\mathbf{X}}^{(i)\top} \mathbf{Y}_0 \,/\, n$
    \State $\mathbf{w}_i \gets$ leading left singular vector of $\mathbf{S}_{XY}^{(i)}$
    \If{sparse} \replaced{$\mathcal{A}_x \gets \mathcal{A}_x \cup \{k : |w_{ik}| > \eta_x \max_l |w_{il}|\}$; \ $w_{ik} \gets 0$ for $k \notin \mathcal{A}_x$}{$\mathbf{w}_i \gets \operatorname{sign}(\mathbf{w}_i) \odot \max\bigl(|\mathbf{w}_i| - \eta_x \max|\mathbf{w}_i|,\, 0\bigr)$} \EndIf
    \State $\mathbf{t}_i \gets \tilde{\mathbf{X}}^{(i)} \mathbf{w}_i$; \quad $\mathbf{p}_i \gets \tilde{\mathbf{X}}^{(i)\top} \mathbf{t}_i \,/\, (\mathbf{t}_i^\top \mathbf{t}_i)$
    \State $\tilde{\mathbf{X}}^{(i+1)} \gets \tilde{\mathbf{X}}^{(i)} - \mathbf{t}_i \mathbf{p}_i^\top$
\EndFor
\Statex \textit{--- $\mathbf{Y}$-block deflation ---}
\For{$j = 1, \ldots, h_y$}
    \State $\mathbf{S}_{YX}^{(j)} \gets \tilde{\mathbf{Y}}^{(j)\top} \mathbf{X}_0 \,/\, n$
    \State $\mathbf{v}_j \gets$ leading left singular vector of $\mathbf{S}_{YX}^{(j)}$
    \If{sparse} \replaced{$\mathcal{A}_y \gets \mathcal{A}_y \cup \{k : |v_{jk}| > \eta_y \max_l |v_{jl}|\}$; \ $v_{jk} \gets 0$ for $k \notin \mathcal{A}_y$}{$\mathbf{v}_j \gets \operatorname{sign}(\mathbf{v}_j) \odot \max\bigl(|\mathbf{v}_j| - \eta_y \max|\mathbf{v}_j|,\, 0\bigr)$} \EndIf
    \State $\mathbf{u}_j \gets \tilde{\mathbf{Y}}^{(j)} \mathbf{v}_j$; \quad $\mathbf{q}_j \gets \tilde{\mathbf{Y}}^{(j)\top} \mathbf{u}_j \,/\, (\mathbf{u}_j^\top \mathbf{u}_j)$
    \State $\tilde{\mathbf{Y}}^{(j+1)} \gets \tilde{\mathbf{Y}}^{(j)} - \mathbf{u}_j \mathbf{q}_j^\top$
\EndFor
\Statex \textit{--- Regression coefficients ---}
\State $\hat{\mathbf{B}}_{\text{scaled}} \gets \replaced[R2.5]{\mathbf{W} (\mathbf{W}^\top \mathbf{X}_0^\top \mathbf{X}_0 \mathbf{W})^{-1} \mathbf{W}^\top \mathbf{X}_0^\top \mathbf{Y}_0 \mathbf{V} \mathbf{V}^\top}{\mathbf{W} (\mathbf{W}^\top \mathbf{X}_0 \mathbf{W})^{-1} (\mathbf{W}^\top \mathbf{X}_0)^\top (\mathbf{X}_0^\top \mathbf{Y}_0 \mathbf{V}) \mathbf{V}^\top}$
\State \replaced[R2.5]{$\hat{\mathbf{B}} \gets \mathbf{S}_X^{-1} \hat{\mathbf{B}}_{\text{scaled}} \mathbf{S}_Y$; \quad $\hat{\mathbf{b}}_0 \gets \mathbf{m}_Y - \hat{\mathbf{B}}^\top \mathbf{m}_X$}{Rescale $\hat{\mathbf{B}}_{\text{scaled}}$ to the original data scale $\to \hat{\mathbf{B}}$}
\State \Return $\hat{\mathbf{B}}$, $\mathbf{W}$, $\mathbf{V}$, $\mathbf{T}$, $\mathbf{U}$, $\mathbf{P}$, $\mathbf{Q}$
\end{algorithmic}
\end{algorithm}

In the sparse variant \citep{serneels2025}, \replaced{variables whose weight is small relative to the largest weight of the current component, i.e.\ $|w_{ik}| \leq \eta_x \max_l |w_{il}|$, are eliminated (steps 7 and 14 in Algorithm~\ref{alg:twoblock}): their weights (and loadings) are set to zero, whereas the weights of the retained variables keep their values. A variable that has been retained for an earlier component remains active for all subsequent components, such that the active sets $\mathcal{A}_x$ and $\mathcal{A}_y$ of selected variables grow with the number of components, as in sparse PLS \citep{chun2010}. This differs slightly from the sparse NIPALS algorithm (SNIPLS) that underlies SPRM \citep{hoffmann2015}, which uses the same selection rule and cumulative active set, but in addition shrinks the retained weights by soft thresholding, i.e.\ $w_{ik} \gets \operatorname{sign}(w_{ik})\bigl(|w_{ik}| - \eta_x \max_l |w_{il}|\bigr)$. In twoblock, the retained weights are not shrunk, which avoids the downward bias that soft thresholding induces in the retained weights. The parameters}{each weight vector is soft-thresholded (steps 7 and 14 in Algorithm~\ref{alg:twoblock}) and} $\eta_x, \eta_y \in [0, 1)$ control the degree of sparsity in the respective blocks. When $\eta_x = \eta_y = 0$, the dense twoblock algorithm is recovered.\deleted{ In the algorithm and elsewhere, $\odot$ denotes the Hadamard (element-wise) matrix product.} \added[R2.5]{In step 1, $\mathbf{m}_X$, $\mathbf{m}_Y$ denote the column locations and $\mathbf{S}_X$, $\mathbf{S}_Y$ the diagonal matrices of column scales used to centre and scale $\mathbf{X}$ and $\mathbf{Y}$ (with $\mathbf{S} = \mathbf{I}$ if no scaling is applied); for classical twoblock, these are the column means and standard deviations, whereas RTB replaces them by robust estimates (Section~\ref{sec:rtb}). Step 20 transforms the coefficients back to the original measurement scale, i.e.\ $\hat{B}_{kl} = \hat{B}_{\text{scaled},kl}\, s_{Y,l} / s_{X,k}$, and computes the corresponding intercept $\hat{\mathbf{b}}_0$, such that predictions for new cases $\mathbf{x}$ are obtained as $\hat{\mathbf{y}} = \hat{\mathbf{b}}_0 + \hat{\mathbf{B}}^\top\mathbf{x}$.}

\added[R1.D]{%
\begin{remark}\label{rem:rgcca}
For two blocks, RGCCA \citep{tenenhaus2011,tenenhaus2017} with shrinkage parameters $\tau_1 = \tau_2 = 1$ maximizes $\mathrm{cov}(\mathbf{X}\mathbf{a}_1, \mathbf{Y}\mathbf{a}_2)$ subject to $\|\mathbf{a}_1\| = \|\mathbf{a}_2\| = 1$, which is Tucker's inter-battery factor analysis \citep{tucker1958}. Its solution is given by the leading left and right singular vectors of $\mathbf{X}_0^\top\mathbf{Y}_0$. In Algorithm~\ref{alg:twoblock}, $\mathbf{w}_1$ is the leading left singular vector of $\mathbf{X}_0^\top\mathbf{Y}_0$ and $\mathbf{v}_1$ is the leading left singular vector of $\mathbf{Y}_0^\top\mathbf{X}_0$, i.e.\ the leading right singular vector of $\mathbf{X}_0^\top\mathbf{Y}_0$. Hence, the first dense twoblock weight vectors in both blocks coincide (up to sign) with the first RGCCA block weight vectors for $\tau_1 = \tau_2 = 1$. The methods differ from the second component onwards: twoblock deflates $\mathbf{X}$ while keeping $\mathbf{Y}_0$ fixed to obtain the $\mathbf{X}$ components, and vice versa for the $\mathbf{Y}$ components, such that both blocks can have a different number of components, whereas RGCCA deflates all blocks and extracts the components of all blocks jointly at each stage. \added{The latter scheme corresponds to Wold's two-block (mode A) PLS algorithm, which \cite{cook2023} (their Section~4.4) show to have notable limitations compared to the separate deflation sequences used by twoblock.} In the sparse case, the thresholding of the singular vectors (steps 7 and 14 of Algorithm~\ref{alg:twoblock}) does not in general yield the same weights as the $\ell_1$ constrained SGCCA solution \citep{tenenhaus2014}.
\end{remark}
}

\subsection{Robust Twoblock (RTB)}\label{sec:rtb}

To robustify the twoblock algorithm, the approach of iterative reweighting is selected. The new method hereby builds upon prior successful application of iterative reweighting in the context of univariate partial least squares, which has lead to the dense \citep{serneels2005} and sparse \citep{hoffmann2015} partial robust M regression estimators. While other paths to robustification could be pursued, the iterative reweighting is elegant for various reasons: it is computationally tractable, has been shown to yield robustness against various types of deviations from normality and it can yield estimators that work in case there are more variables than cases.  In various fields of application, such as chemometrics and bioinformatics, data with either $p > n$, $q > n$, or both, can be encountered. The prerequisite that the method can be computed for such data precludes other plausible approaches, such as plugging subset selection based estimates, such as MCD \citep{rousseeuw1985} or LTS \citep{rousseeuw1984}. Likewise, generalized spatial sign pre-processing \citep{serneels2024}, which originally was a step in the calculation of the generalized spatial sign covariance matrix \citep{raymaekers2019generalized}, cannot be applied to data that have more variables than cases. 

Accounting for the prerequisite of yielding an estimator that is operable when $p>n$ or $q>n$, RTB wraps the twoblock estimation into an iterative reweighting scheme. The key idea is to assign case weights based on outlyingness in the respective score spaces, and to re-estimate the model on the weighted data until convergence. \added[R1.10]{More precisely, let $\boldsymbol{\Omega}_X = \operatorname{diag}(\boldsymbol{\omega}^{(X)})$ and $\boldsymbol{\Omega}_Y = \operatorname{diag}(\boldsymbol{\omega}^{(Y)})$ contain the case weights of the $\mathbf{X}$ and $\mathbf{Y}$ blocks. At each iteration, twoblock is applied to the weighted data $\boldsymbol{\Omega}_X^{1/2}\mathbf{X}_s$ and $\boldsymbol{\Omega}_Y^{1/2}\mathbf{Y}_s$. By~\eqref{eq:tbcrit}, the $i$-th $\mathbf{X}$-block weight vector then solves
\begin{equation}\label{eq:rtbcrit}
    \mathbf{w}_i = \operatorname*{argmax}_{\|\mathbf{w}\| = 1} \sum_{l=1}^{q} \Bigl( \sum_{k=1}^{n} \sqrt{\omega_k^{(X)} \omega_k^{(Y)}}\; \tilde{t}_{k}(\mathbf{w})\, y_{s,kl} \Bigr)^2,
\end{equation}
with $\tilde{t}_k(\mathbf{w})$ the $k$-th entry of $\tilde{\mathbf{X}}_s^{(i)}\mathbf{w}$, i.e.\ a weighted covariance criterion in which each case contributes with weight $\sqrt{\omega_k^{(X)}\omega_k^{(Y)}}$; the $\mathbf{Y}$-block weights solve the analogous problem. The case weights are in turn functions of the score distances of the current solution, $\boldsymbol{\omega}^{(X)} = w_H\bigl(\mathbf{d}(\mathbf{T})\bigr)$ and $\boldsymbol{\omega}^{(Y)} = w_H\bigl(\mathbf{d}(\mathbf{U})\bigr)$, with the weight function $w_H$ defined in~\eqref{eq:hampel} below. RTB thus computes a fixed point of the weighted twoblock estimating equations~\eqref{eq:rtbcrit}, with weights of the form $w(d) = \psi(d)/d$ derived from the $\psi$-function of an M-estimator. This is the same construction as in PRM and SPRM \citep{serneels2005,hoffmann2015}, and it is the two-block analogue of the iteratively reweighted least squares algorithm, which computes M-estimates as fixed points of weighted least squares problems \citep{maronna2019robust}. Because the twoblock criterion is a covariance maximization rather than a loss minimization, RTB is defined through this fixed point rather than as the minimizer of a single closed-form robust loss.}

\paragraph{Initialization.}
In the context of iterative reweighting and M-estimation, it is well known that the eventual robustness of the estimates strongly depends on the selection of the starting values: reweighted methods can still break down in case the starting values themselves break down in the presence of outliers. This begins with the centring and scaling of the data. Ideally, data are centred about a highly robust estimate of location that also lies within the convex hull of the data. Likewise, a highly robust scale estimator should be chosen. Good selections for these two preprocessing steps are the $\ell_1$ median location estimator (for computational aspects thereof, see e.g. \cite{fritz2012comparison}) and the $\tau_2$ scale estimator \citep{maronna2002}, but other options are possible. If computational efficiency is paramount, then the column-wise median and median absolute deviation could be opted for. 

Another important choice to make for iterative reweighting is the weight function. Continuous reweighting functions, such as the Fair function, can be adopted. However, the Fair function will downweight all cases outside the exact location of the distribution, which yields weights that can be difficult to interpret and may have unsatisfactory predictive performance due to excessive downweighting of non-outlying cases. Discontinuous downweighting functions, that use a certain cutoff beyond which cases are considered outliers and downweighting starts, are often seen as options that are easy to implement, that yield highly interpretable caseweights, and also lead to good statistical robustness. Popular discontinuous downweighting functions are the Huber and Hampel functions. The latter is applied throughout the simulation and example section of this paper. The Hampel function is defined as: 
\begin{equation}\label{eq:hampel}
    w_H(d_i) =
    \begin{cases}
        1 & \text{if } d_i \leq c_1, \\[4pt]
        \dfrac{c_1}{d_i} & \text{if } c_1 < d_i \leq c_2, \\[6pt]
        \dfrac{c_1\,(c_3 - d_i)}{d_i\,(c_3 - c_2)} & \text{if } c_2 < d_i \leq c_3, \\[6pt]
        0 & \text{if } d_i > c_3,
    \end{cases}
\end{equation}
\replaced[R2.6]{where $d_i \geq 0$ is the distance of case $i$ in the respective score space, as defined in Algorithms~\ref{alg:startweights} and~\ref{alg:rtb}, and $0 < c_1 < c_2 < c_3$ are three cutoff values: cases with $d_i \leq c_1$ receive full weight, downweighting starts beyond $c_1$, becomes steeper beyond $c_2$, and cases with $d_i > c_3$ receive weight zero. Equation~\eqref{eq:hampel} is the weight function $w_H(d) = \psi_H(d)/d$ that corresponds to Hampel's three-part redescending $\psi$-function $\psi_H$ \citep{maronna2019robust}.}{where the case $d_i \in \mathbb{R}^{+}$ can be any positive scalar, but in what follows, is typically computed from the data as some distance metric. The Hampel function depends on three cutoff values, which increase the harshness of downweighting depending on how eccentric the corresponding case is.} These cutoff values are tunable parameters of the algorithm\replaced[R2.6, R2.11]{. They are specified through three probabilities $p_1 < p_2 < p_3$ and set to the corresponding quantiles of the $\chi^2$ distribution, $c_k = \chi^2_{h, p_k}$ ($k = 1, 2, 3$), where the degrees of freedom $h$ equal the number of components in the respective block ($h_x$ for $\boldsymbol{\omega}^{(X)}$, $h_y$ for $\boldsymbol{\omega}^{(Y)}$, and $r$ for the starting weights in Algorithm~\ref{alg:startweights}).}{ and they can be set according to corresponding quantiles, for instance quantiles of the $\chi^2$ distribution for squared distances.} This approach has been successfully adopted before in related contexts, for instance in SPRM regression or in cellwise robust M regression \citep{FILZMOSER2020106944}.     

The starting weight computation is formalized in Algorithm~\ref{alg:startweights}. It is applied independently to both the $\mathbf{X}$ and $\mathbf{Y}$ blocks. \added[R2.8]{To avoid confusion with the weight vectors $\mathbf{w}_i$ and $\mathbf{v}_j$ of Algorithm~\ref{alg:twoblock}, case weights are denoted $\boldsymbol{\omega}$ throughout. The starting distances are computed in a low-dimensional principal component space of the robustly scaled block, whose dimension $r$ is chosen by the broken-stick rule; this also makes the starting weights computable when the block has more variables than cases.}

\begin{algorithm}[t]
\caption{Starting Weights}\label{alg:startweights}
\begin{algorithmic}[1]
\Require \replaced[R2.8]{Robustly centred and scaled data block $\mathbf{Z}_s \in \mathbb{R}^{n \times m}$ (i.e.\ $\mathbf{X}_s$ with $m = p$, or $\mathbf{Y}_s$ with $m = q$), weight function $w_H$}{Scaled data block $\mathbf{Z}_s \in \mathbb{R}^{n \times d}$, weighting function $\psi$}
\State \added[R2.8]{$\mathbf{Z}_s = \mathbf{L}\mathbf{S}\mathbf{R}^\top$} \Comment{\added{Singular value decomposition}}
\State \added[R2.8]{$r \gets$ number of components retained by the broken-stick rule}
\State \added[R2.8]{$\tilde{\mathbf{L}} \gets$ first $r$ columns of $\mathbf{L}$, robustly centred and scaled per column}
\State \replaced[R2.8]{$d_i \gets \|\tilde{\boldsymbol{\ell}}_i\|_2 \,/\, \operatorname{median}_k \|\tilde{\boldsymbol{\ell}}_k\|_2$, \ $i = 1, \ldots, n$, with $\tilde{\boldsymbol{\ell}}_i$ the $i$-th row of $\tilde{\mathbf{L}}$}{$\mathbf{d} \gets \bigl[\|\mathbf{z}_{r,i}\|_2 \,/\, \operatorname{median}_i \|\mathbf{z}_{r,i}\|_2\bigr]_{i=1}^{n}$} \Comment{Standardized casewise distances}
\State \replaced[R2.8]{$\omega_i \gets w_H(d_i)$, with cutoffs $c_k = \chi^2_{r, p_k}$}{$\mathbf{w} \gets \psi(\mathbf{d})$} \Comment{Apply weight function}
\State \replaced[R2.8]{$\omega_i \gets \max(\omega_i,\, 10^{-6})$}{$w_i \gets \max(w_i,\, 10^{-6})$} for all $i$ \Comment{Floor small weights}
\State \Return \replaced[R2.8]{$\boldsymbol{\omega}$}{$\mathbf{w}$}
\end{algorithmic}
\end{algorithm}

The RTB initialization then proceeds as:
\begin{enumerate}
    \item Robustly centre and scale $\mathbf{X} \to \mathbf{X}_s$ and $\mathbf{Y} \to \mathbf{Y}_s$\added[R2.7]{, by default with the column-wise median and median absolute deviation (MAD); the $\ell_1$-median, the $\tau$-scale or the k-step LTS location and scale can be selected instead}.
    \item \replaced[R2.8]{$\boldsymbol{\omega}^{(X)} \gets \textsc{Starting Weights}(\mathbf{X}_s, w_H)$}{$\mathbf{w}^{(X)} \gets \textsc{Starting Weights}(\mathbf{X}_s, \psi)$}; \replaced[R2.8]{$\boldsymbol{\omega}^{(Y)} \gets \textsc{Starting Weights}(\mathbf{Y}_s, w_H)$}{$\mathbf{w}^{(Y)} \gets \textsc{Starting Weights}(\mathbf{Y}_s, \psi)$}.
    \item Form weighted data: \replaced[R2.8]{$\mathbf{X}_w = \boldsymbol{\Omega}_X^{1/2} \mathbf{X}_s$ and $\mathbf{Y}_w = \boldsymbol{\Omega}_Y^{1/2} \mathbf{Y}_s$}{$\mathbf{X}_w = \operatorname{diag}(\sqrt{\mathbf{w}^{(X)}}) \, \mathbf{X}_s$ and $\mathbf{Y}_w = \operatorname{diag}(\sqrt{\mathbf{w}^{(Y)}}) \, \mathbf{Y}_s$}.
\end{enumerate}

\paragraph{Iterative reweighting.}
Repeat until convergence (or a maximum number of iterations):
\begin{enumerate}
    \item Fit a twoblock model (dense or sparse) to $\mathbf{X}_w$ and $\mathbf{Y}_w$ with centring set to `mean' and scaling set to `None' (since the data are already robustly preprocessed).
    \item Extract the $\mathbf{X}$-block scores $\mathbf{T}$ and $\mathbf{Y}$-block scores $\mathbf{U}$ from the fitted model, unweighting them by dividing \replaced[R2.9]{each row by the square root of the corresponding case weight}{by $\sqrt{\mathbf{w}}$}.
    \item Update $\mathbf{X}$-block weights: \replaced[R2.4]{robustly centre and scale each column of the unweighted scores, compute for each case the Euclidean norm $\delta_i$ of its standardized score vector, set $d_i = \delta_i / \operatorname{median}_k \delta_k$ and apply the weight function $w_H$ to obtain updated $\boldsymbol{\omega}^{(X)}$}{compute robust Mahalanobis distances of the unweighted scores, apply the weighting function $\psi$ to obtain updated $\mathbf{w}^{(X)}$}.
    \item Update $\mathbf{Y}$-block weights analogously from the $\mathbf{Y}$ scores to obtain updated \replaced[R2.8]{$\boldsymbol{\omega}^{(Y)}$}{$\mathbf{w}^{(Y)}$}.
    \item Re-form $\mathbf{X}_w$ and $\mathbf{Y}_w$ with the updated weights.
    \item Check convergence: \replaced[R2.10]{$\bigl|\|\hat{\mathbf{B}}^{(t)}_{s}\|_F^2 - \|\hat{\mathbf{B}}^{(t-1)}_{s}\|_F^2\bigr| / \|\hat{\mathbf{B}}^{(t-1)}_{s}\|_F^2 < \tau$, where $\hat{\mathbf{B}}^{(t)}_{s}$ is the (scaled) coefficient matrix at iteration $t$ (by default $\tau = 0.01$, with at most 100 iterations)}{$|\|\mathbf{b}^{(t)}\|^2 - \|\mathbf{b}^{(t-1)}\|^2| / \|\mathbf{b}^{(t-1)}\|^2 < \tau$}.
\end{enumerate}

\paragraph{Final estimates.}
After convergence, the regression coefficients are rescaled to the original data scale\added[R2.5]{ as in step 20 of Algorithm~\ref{alg:twoblock}, with the robust column scales; the intercept is computed as the column-wise median of $\mathbf{Y} - \mathbf{X}\hat{\mathbf{B}}$}.
The combined case weight for observation $i$ is \replaced[R2.8]{$\omega_i = \omega_i^{(X)} \cdot \omega_i^{(Y)}$. It serves as a diagnostic (Section~\ref{sec:examples}); in the estimation, each case enters the cross-covariance through $\sqrt{\omega_i^{(X)}\omega_i^{(Y)}}$, as shown in~\eqref{eq:rtbcrit}.}{$w_i = w_i^{(X)} \cdot w_i^{(Y)}$.}

\added[R2.4]{The distances $d_i$ are computed from robustly standardized scores. This is a Mahalanobis-type distance in the score space: since the components extracted within a block are mutually orthogonal, standardizing each score column by its (robust) scale is, up to the robust estimation of location and scale, equivalent to computing a Mahalanobis distance of the scores. A Euclidean distance on the unstandardized scores would be dominated by the first components, which capture most variance, such that cases that are outlying along a later component would go undetected. The variance explained by each component thus deliberately does not enter the distance, as for score distances in robust principal component analysis. Only the $h_x$ (or $h_y$) components retained in the model enter the distances, and these numbers are selected by cross-validation (Section~\ref{sec:examples}).}

The robust twoblock algorithm is summarized in Algorithm~\ref{alg:rtb}.

\begin{algorithm}[t]
\caption{Robust Twoblock (RTB)}\label{alg:rtb}
\begin{algorithmic}[1]
\Require $\mathbf{X} \in \mathbb{R}^{n \times p}$, $\mathbf{Y} \in \mathbb{R}^{n \times q}$, components $h_x$, $h_y$, \replaced[R2.8]{weight function $w_H$}{weighting function $\psi$}
\State \replaced{$\mathbf{m}_X, \mathbf{m}_Y \gets$ robust column locations of $\mathbf{X}$, $\mathbf{Y}$ (e.g.\ medians); \ $\mathbf{S}_X, \mathbf{S}_Y \gets$ diagonal matrices of robust column scales (e.g.\ MAD; $\mathbf{I}$ if unscaled); \ $\mathbf{X}_s \gets (\mathbf{X} - \mathbf{1}_n \mathbf{m}_X^\top)\mathbf{S}_X^{-1}$, \ $\mathbf{Y}_s \gets (\mathbf{Y} - \mathbf{1}_n \mathbf{m}_Y^\top)\mathbf{S}_Y^{-1}$}{Robustly centre and scale $\mathbf{X} \to \mathbf{X}_s$, $\mathbf{Y} \to \mathbf{Y}_s$}
\State $\boldsymbol{\omega}^{(X)} \gets$ \Call{Starting Weights}{$\mathbf{X}_s$, $w_H$}
\State $\boldsymbol{\omega}^{(Y)} \gets$ \Call{Starting Weights}{$\mathbf{Y}_s$, $w_H$}
\State $\mathbf{X}_w \gets \boldsymbol{\Omega}_X^{1/2} \mathbf{X}_s$; \quad $\mathbf{Y}_w \gets \boldsymbol{\Omega}_Y^{1/2} \mathbf{Y}_s$
\Repeat
    \State Fit twoblock$(\mathbf{X}_w, \mathbf{Y}_w)$ $\to$ scores $\mathbf{T}$, $\mathbf{U}$, \replaced[R2.10]{coefficient matrix $\hat{\mathbf{B}}_s$}{coefficients $\mathbf{b}$}
    \State $\mathbf{T}_{\text{uw}} \gets \boldsymbol{\Omega}_X^{-1/2}\mathbf{T}$; \quad $\mathbf{U}_{\text{uw}} \gets \boldsymbol{\Omega}_Y^{-1/2}\mathbf{U}$ \Comment{\added[R2.9]{unweight the scores}}
    \State $\boldsymbol{\omega}^{(X)} \gets w_H\bigl(\mathbf{d}(\mathbf{T}_{\text{uw}})\bigr)$; \quad $\boldsymbol{\omega}^{(Y)} \gets w_H\bigl(\mathbf{d}(\mathbf{U}_{\text{uw}})\bigr)$
    \State $\mathbf{X}_w \gets \boldsymbol{\Omega}_X^{1/2} \mathbf{X}_s$; \quad $\mathbf{Y}_w \gets \boldsymbol{\Omega}_Y^{1/2} \mathbf{Y}_s$
\Until{\replaced[R2.10]{$\|\hat{\mathbf{B}}_s\|_F^2$ converges}{$\|\mathbf{b}\|^2$ converges}}
\State Rescale coefficients to original scale\added[R2.5]{ and compute the intercept}
\State \Return $\hat{\mathbf{B}}$, \added[R2.5]{$\hat{\mathbf{b}}_0$,} $\boldsymbol{\omega}^{(X)}$, $\boldsymbol{\omega}^{(Y)}$
\end{algorithmic}
\end{algorithm}

\subsection{Properties}
As the simulation study in the ensuing section will show, RTB inherits the dimension reduction and variable selection properties of twoblock (dense or sparse), while gaining robustness to outliers. The independent weighting of the $\mathbf{X}$ and $\mathbf{Y}$ blocks allows the method to handle leverage points (outliers in $\mathbf{X}$ are downweighted via $\boldsymbol{\omega}^{(X)}$), vertical outliers (outliers in $\mathbf{Y}$ are downweighted via $\boldsymbol{\omega}^{(Y)}$) and joint contamination, since both weights \replaced[R1.10]{enter the weighted cross-covariance in~\eqref{eq:rtbcrit}}{contribute to the combined weight $\mathbf{w} = \mathbf{w}^{(X)} \odot \mathbf{w}^{(Y)}$}.

On clean data, the Hampel weighting function with standard cutoffs given as distributional quantiles ($p_1 = 0.95$, $p_2 = 0.975$, $p_3 = 0.999$) assigns weight 1 to the vast majority of observations, so RTB converges to a solution close to classical twoblock. More aggressive cutoffs (e.g.\ $p_1 = 0.75$, $p_2 = 0.90$, $p_3 = 0.95$) can be used when heavier contamination is expected. The latter option will be used in the results presented hereinafter. \added[R1.11]{In applications, the hyperparameters of RTB are the numbers of components $h_x$ and $h_y$, for sparse RTB also the sparsity parameters $\eta_x$ and $\eta_y$, and the Hampel probabilities $p_1, p_2, p_3$. As illustrated in Section~\ref{sec:examples}, $h_x$, $h_y$, $\eta_x$ and $\eta_y$ are selected by cross-validation, since no prior knowledge of the true structure is available. The Hampel probabilities can be included in the cross-validation grid as well. If they are not, the two settings above can serve as a guideline: the default probabilities yield a highly efficient estimator that only discards gross outliers, whereas the more aggressive setting is preferable when a substantial fraction of contamination is expected. The case weights of the fitted model provide a diagnostic for that choice: if many cases are downweighted under the default setting, the more aggressive setting is advisable.} \added[R2.16]{When the data are contaminated, the validation folds are contaminated as well. The cross-validation criterion should then be robust, e.g.\ a trimmed prediction error or the prediction error computed on the cases with full weight only, since otherwise the large residuals of the outliers dominate the criterion (see Section~\ref{sec:gas}).}

\section{Simulation study}\label{sec:simulation}

In this simulation study, the robustness and sparsity properties of the method will be assessed. The simulations are set up such that they generate data according to a latent variable model with known \replaced[R1.11]{dimension}{model complexity} and are thus apt to be modeled by each of the twoblock methods without the need for tuning the corresponding parameters. Outliers are added in six different scenarios, i.e. at two different proportions, and either as vertical outliers only, as leverage points only, or as both types of outliers combined. To assess the variable selection properties of the sparse method, noise variables will be added to the $\mathbf{X}$ block in two proportions, with one scenario adding far fewer noise variables than informative variables, whereas the other scenario adds {\em more} uninformative variables than informative ones. Finally, dimensions of the $\mathbf{X}$ block data will also be evaluated both in a scenario where $n < p$ and where $p > n$. Results for each setting are reported as averages over two hundred runs. In total, the simulation study thereby evaluates 42 scenarios\added[R1.12]{ (six dimensionality settings, each with seven contamination settings), i.e.\ 8{,}400 simulated data sets, each of which is analysed with four methods, for} a grand total of \replaced[R1.12]{33{,}600 model fits}{33200 individual runs}.    

\subsection{Design}

Data are generated from a true latent variable model\added[R1.13]{, in which a common set of unobserved factors $\mathbf{Z}$ generates both blocks}:
\begin{align}
    \mathbf{X} &= \replaced[R1.13]{\mathbf{Z} \boldsymbol{\Gamma}^\top + \mathbf{N}_X, \quad \mathbf{Z} \in \mathbb{R}^{n \times k},\; \boldsymbol{\Gamma} \in \mathbb{R}^{p_{\text{signal}} \times k},\; (\mathbf{N}_X)_{ij}}{\mathbf{T} \mathbf{P}^\top + \mathbf{E}, \quad \mathbf{T} \in \mathbb{R}^{n \times k},\; \mathbf{P} \in \mathbb{R}^{p_{\text{signal}} \times k},\; E_{ij}} \sim \mathcal{N}(0, \sigma_e^2), \\
    \mathbf{Y} &= \replaced[R1.13]{\mathbf{Z} \mathbf{C} + \mathbf{N}_Y, \quad \mathbf{C} \in \mathbb{R}^{k \times q},\; (\mathbf{N}_Y)_{ij}}{\mathbf{T} \mathbf{C} + \mathbf{F}, \quad \mathbf{C} \in \mathbb{R}^{k \times q},\; F_{ij}} \sim \mathcal{N}(0, \sigma_f^2),
\end{align}
where \replaced[R1.13]{$\boldsymbol{\Gamma}$ has orthonormal columns, $Z_{ij} \sim \mathcal{N}(0,1)$}{$\mathbf{P}$ has orthonormal columns, $T_{ij} \sim \mathcal{N}(0,1)$}, and $C_{ij} \sim \mathcal{N}(0,1)$.
The true regression coefficient is \replaced[R1.13]{$\mathbf{B}_{\text{true}} = [\boldsymbol{\Gamma} \mathbf{C};\; \mathbf{0}]$}{$\mathbf{B}_{\text{true}} = [\mathbf{P} \mathbf{C};\; \mathbf{0}]$}, padded with zeros for uninformative variables. \added[R1.4, R1.13, R2.12]{Here, distinct symbols are used for the data generating quantities, to distinguish them from the scores $\mathbf{T} = \mathbf{X}\mathbf{W}$ and $\mathbf{U} = \mathbf{Y}\mathbf{V}$ and the residuals $\mathbf{E}$ and $\mathbf{F}$ that are estimated by the methods. The factors $\mathbf{Z}$ are latent variables in the classical (reflective) sense, which generate the observed variables in both blocks; the model is thus symmetric in $\mathbf{X}$ and $\mathbf{Y}$, in that neither block is generated from the other. The estimated scores, in contrast, are components, i.e.\ linear combinations of the observed variables.}
Uninformative variables are appended as columns of $\mathcal{N}(0, \sigma_e^2)$ noise. \added[R1.14]{The true dimension of the latent structure is $k$: the part of $\mathbf{X}$ that is relevant to $\mathbf{Y}$ and the part of $\mathbf{Y}$ that is relevant to $\mathbf{X}$ are both of dimension $k$ (since $q > k$, $\mathbf{Y}$ additionally contains a direction that carries noise only). The correct numbers of components in Algorithm~\ref{alg:twoblock} are therefore $h_x = k$ and $h_y = \min(k, q) = k$.}

We set $n = 100$, $k = 3$, $q = 4$, $\sigma_e = \sigma_f = 0.5$, and vary:
\begin{itemize}
    \item \textbf{Dimensionality}: $p_{\text{signal}} = 20$ (with $p_{\text{noise}} \in \{0, 10, 80\}$) and $p_{\text{signal}} = 150$ (with $p_{\text{noise}} \in \{0, 50, 250\}$), covering both $p < n$ and $p > n$ regimes.
    \item \textbf{Contamination}: $0\%$, $10\%$, or $20\%$ of cases, with outliers in $\mathbf{X}$ only, $\mathbf{Y}$ only, or both, introduced as additive shifts of magnitude 10\added[R2.12]{ with a random sign, applied to every variable of the contaminated block(s) for the selected cases}.
\end{itemize}
\added[R1.15]{In addition, the complete design is repeated with $q = 2 < k$ (see the end of Section~\ref{sec:simulation}).}

Four methods are compared, each using the true $k$ as the number of $\mathbf{X}$ components\added[R1.14]{ ($h_x = k$)} and $\min(k, q)$ as the number of $\mathbf{Y}$ components\added[R1.14]{ ($h_y = \min(k, q)$)}: dense and sparse classical twoblock (with standard scaling), along with dense and sparse robust twoblock (RTB) with Hampel weighting, median/MAD preprocessing standard cutoffs and the set of $\{p_1 = 0.75, p_2 = 0.90, p_3 = 0.95\}$ for the Hampel function cutoffs. At this point, we note that a more stringent setting for the downweighting cutoffs, such as $\{p_1 = 0.95, p_2 = 0.99, p_3 = 0.999\}$ would yield a less robust, yet highly efficient estimator. However, simulation results for the estimator shown are already quite convincing in terms of efficiency, so results for the more stringent set of parameters will not be presented.  

To evaluate the accuracy of the regression coefficients, the Frobenius norm of the deviation of estimated versus actual regression coefficients $\text{MSE}(\mathbf{B}) = \frac{1}{pq} \|\hat{\mathbf{B}} - \mathbf{B}_{\text{true}}\|_F^2$, averaged over 200 independent repeats, is reported. \added{Note that $\mathbf{B}_{\text{true}}$ is the coefficient matrix of the data generating model, which differs from the population least squares coefficient $\boldsymbol{\Sigma}_{\mathbf{X}}^{-1}\boldsymbol{\Sigma}_{\mathbf{X}\mathbf{Y}} = \mathbf{B}_{\text{true}}/(1+\sigma_e^2)$ by a constant factor; since all methods are evaluated against the same target, this does not affect their comparison.} \added[R1.16]{To evaluate predictive performance, an independent, uncontaminated test set of $n_{\text{test}} = 1000$ cases is generated from the same model (i.e.\ with the same $\boldsymbol{\Gamma}$ and $\mathbf{C}$) for each repeat, and the mean squared prediction error $\text{MSEP} = \frac{1}{n_{\text{test}}\, q}\sum_{i=1}^{n_{\text{test}}} \|\mathbf{y}_i - \hat{\mathbf{y}}_i\|^2$ is reported. For reference, the optimal (population) prediction error in this design is $\sigma_f^2 + \frac{\sigma_e^2}{1+\sigma_e^2}\,\frac{1}{q}\|\mathbf{C}\|_F^2 \approx 0.85$, the second term reflecting that the latent factors are only observed through $\mathbf{X}$ with noise.} To assess variable selection performance of the sparse methods, the $F_1$ score for variable selection is reported. \added[R1.A]{A predictor is considered selected when its row of $\hat{\mathbf{B}}$ contains at least one non-zero entry. With TP, FP and FN the numbers of correctly selected signal variables, selected noise variables and non-selected signal variables, respectively, precision $= \mathrm{TP}/(\mathrm{TP}+\mathrm{FP})$, recall $= \mathrm{TP}/(\mathrm{TP}+\mathrm{FN})$ and $F_1 = 2\,\text{precision} \cdot \text{recall} / (\text{precision} + \text{recall})$, the harmonic mean of both \citep{vanrijsbergen1979}. $F_1 = 1$ corresponds to perfect variable selection.} \added[R1.17]{For both sparse methods, the sparsity parameters are fixed at $\eta_x = 0.5$ and $\eta_y = 0$ throughout the simulation study, rather than tuned by cross-validation in each replicate. Since the same value is used for sparse twoblock and sparse RTB, this does not favour either method, and it isolates the effect of the robust reweighting on estimation and variable selection.} 

\subsection{Results}

\begin{figure}[ht]
\centering
\includegraphics[width=\textwidth]{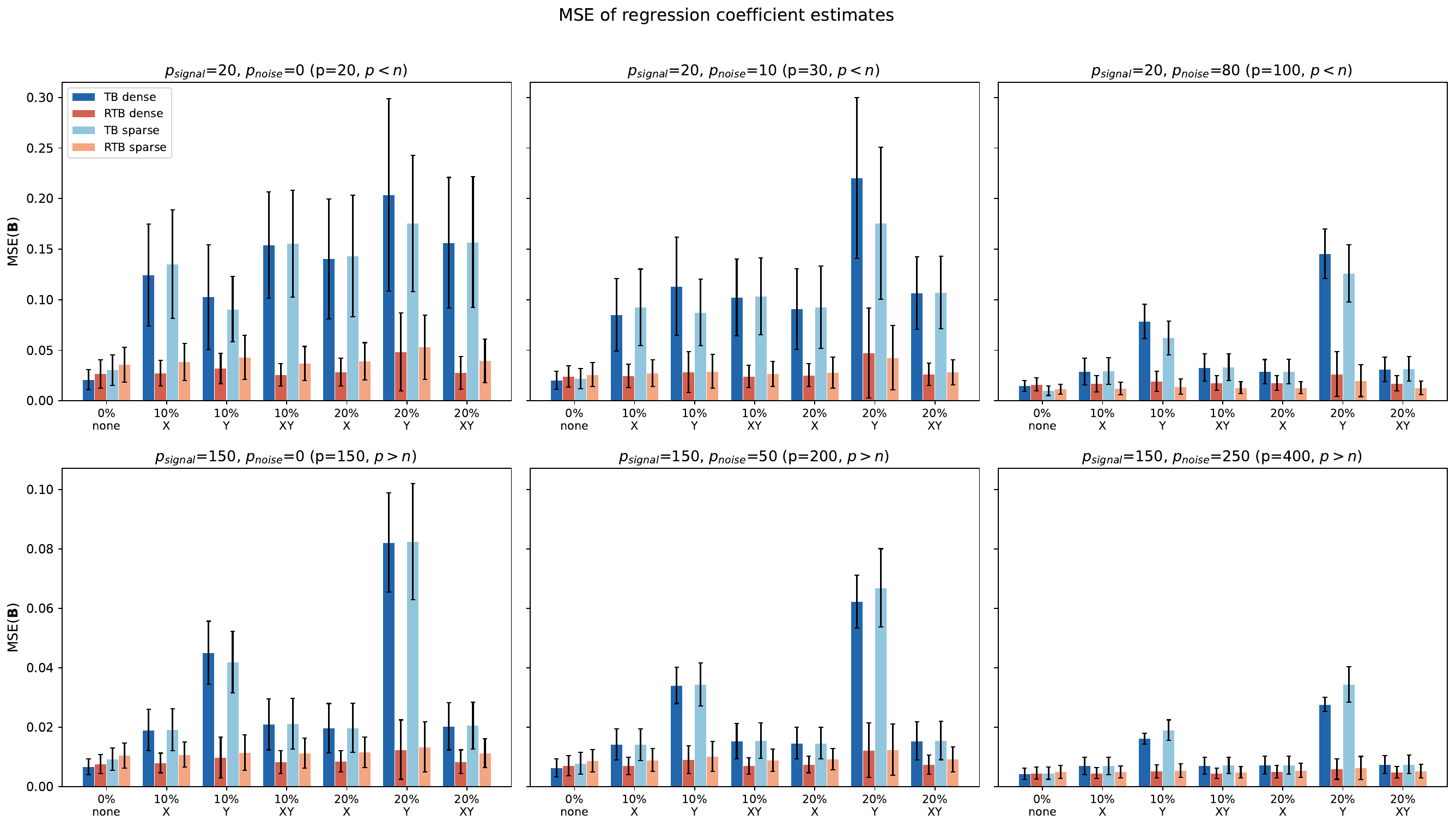}
\caption{MSE of regression coefficient estimates across 42 simulation scenarios (200 repeats each). Dark blue: TB dense; dark red: RTB dense; light blue: TB sparse; salmon: RTB sparse.
Top row: $p \leq n$; bottom row: $p > n$.
Within each panel, the x-axis shows the contamination proportion and type.
Error bars indicate one standard deviation.}
\label{fig:sim_mse}
\end{figure}

\paragraph{Coefficient estimation accuracy.}
Figure~\ref{fig:sim_mse} shows the MSE of the coefficient estimates for all 42 scenarios, arranged in a $2 \times 3$ grid. All four methods---dense and sparse twoblock, dense and sparse RTB---are shown side by side.
The top row corresponds to $p_{\text{signal}} = 20$ ($p \leq n$) and the bottom row to $p_{\text{signal}} = 150$ ($p > n$), with the number of uninformative noise variables increasing from left to right.

\begin{figure}[ht]
\centering
\includegraphics[width=\textwidth]{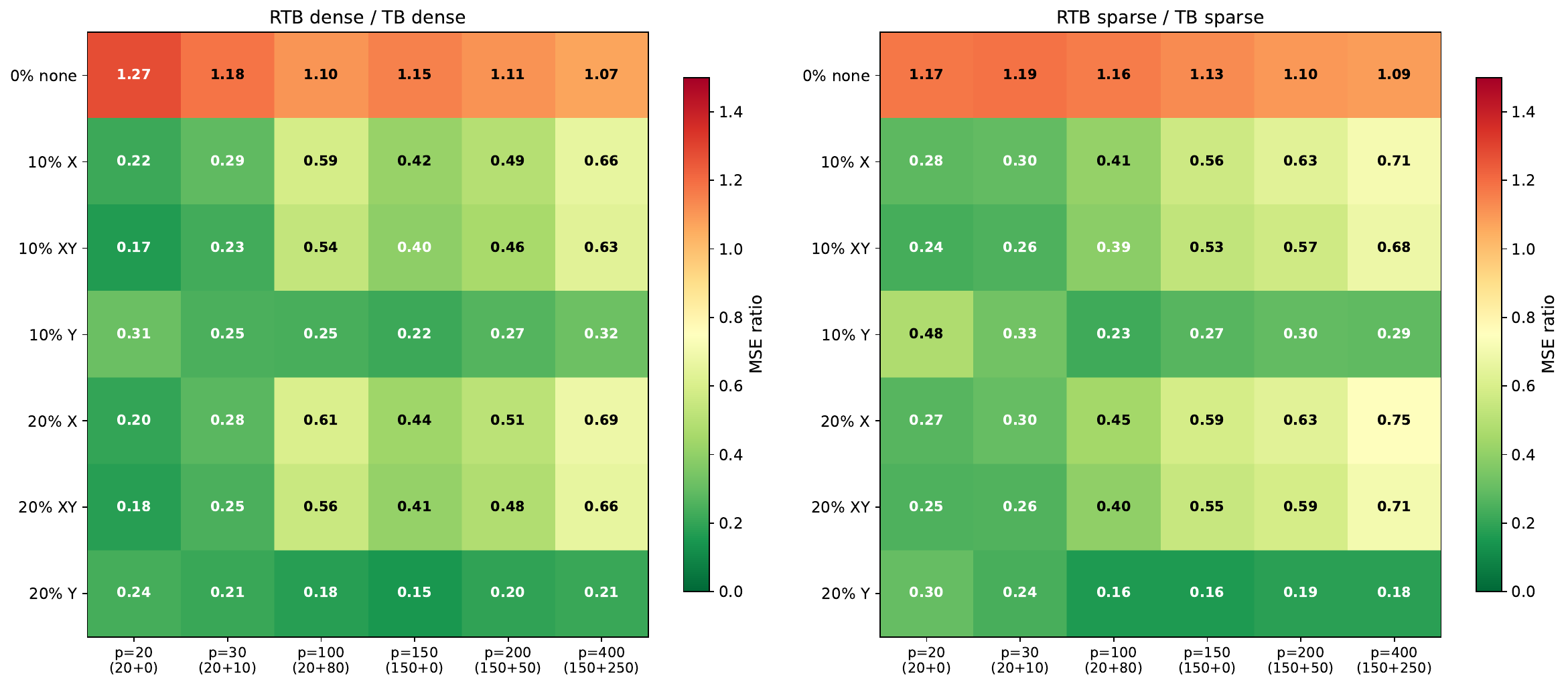}
\caption{MSE ratio of RTB to twoblock for dense (left) and sparse (right) variants across all simulation scenarios.
Green cells ($< 1$) indicate RTB outperforms twoblock; red cells ($> 1$) the reverse.
The only red cells correspond to the uncontaminated baseline.}
\label{fig:sim_heatmap}
\end{figure}

Several patterns emerge consistently across all dimensionality settings:

\begin{enumerate}
    \item \textbf{Clean data.} On uncontaminated data (0\%, none), all four methods perform similarly, with RTB incurring only a modest efficiency loss (\replaced[R2.13]{ratio of the MSE of RTB to the MSE of the corresponding twoblock variant: dense $\approx 1.1$--$1.3$; sparse $\approx 1.1$--$1.2$)}{dense ratio $\approx 1.1$--$1.3$; sparse ratio $\approx 1.1$--$1.2$) relative to the corresponding classical twoblock variant}.

    \item \textbf{$\mathbf{X}$ outliers.} Leverage points in $\mathbf{X}$ cause substantial MSE inflation for both dense and sparse twoblock (e.g.\ dense MSE jumps from 0.021 to 0.124 at 10\% contamination with $p = 20$; sparse from 0.030 to 0.135). Both RTB variants are largely unaffected, with MSE remaining close to the clean-data baseline.

    \item \textbf{$\mathbf{Y}$ outliers.} Vertical outliers in $\mathbf{Y}$ are the most challenging scenario for all methods. RTB reduces MSE substantially relative to classical twoblock (e.g.\ dense ratio 0.31 at 10\% for $p = 20$; sparse ratio 0.48). With majority noise variables ($p_{\text{noise}} = 80$), the sparse RTB achieves particularly strong protection (MSE 0.014 vs 0.062 for sparse twoblock at 10\%).

    \item \textbf{Joint $\mathbf{X}\mathbf{Y}$ outliers.} When both blocks are contaminated, the MSE ratios are similar to the $\mathbf{X}$-only case, since the $\mathbf{X}$-block weights dominate the reweighting. Both dense and sparse RTB maintain their advantage.

    \item \textbf{High-dimensional setting.} The RTB advantage persists in the $p > n$ regime, with similar relative gains. For $p = 400$ ($p_{\text{signal}} = 150$, $p_{\text{noise}} = 250$), RTB dense achieves \added[R2.13]{MSE} ratios \added[R2.13]{(RTB over twoblock, see Figure~\ref{fig:sim_heatmap})} of 0.64--0.66 under $\mathbf{X}$ contamination, and the sparse variant performs comparably.

    \item \textbf{Noise variables.} Adding uninformative variables does not qualitatively change the relative performance of the methods, though absolute MSE values decrease (due to the normalization by $pq$). Sparse methods tend to achieve lower MSE than their dense counterparts when many noise variables are present (e.g.\ $p_{\text{noise}} = 80$: sparse TB 0.010 vs dense TB 0.015 on clean data), reflecting the benefit of variable selection.
\end{enumerate}

Figure~\ref{fig:sim_heatmap} summarizes the relative efficiency of dense and sparse RTB versus the corresponding twoblock variant across all scenarios as heatmaps of MSE ratios.
Values below 1 (green) indicate that RTB outperforms twoblock; values above 1 (red) indicate the reverse.
The only scenarios where twoblock wins are the clean-data cases (ratio $\approx 1.07$--$1.27$), reflecting the modest robustness tax.
Under any form of contamination, RTB consistently achieves ratios well below 1, with the strongest gains for $\mathbf{Y}$ outliers at 20\% contamination. The pattern is highly consistent across the dense and sparse variants.

\added[R1.16]{%
\paragraph{Prediction accuracy.}
Figure~\ref{fig:sim_pred} shows the mean squared prediction error on the uncontaminated test sets. The pattern mirrors that of the coefficient estimates, but the differences between the methods are even more pronounced. On clean data, RTB incurs a small loss in predictive efficiency, with MSEP ratios of RTB to the corresponding twoblock variant of 1.05--1.07 (dense) and 1.04--1.08 (sparse). Under any form of contamination, the prediction error of classical twoblock increases two- to fivefold, whereas that of RTB remains close to its clean-data level. Across all 36 contaminated scenarios, the MSEP ratios of RTB to twoblock range from 0.30 to 0.60 for the dense and from 0.33 to 0.68 for the sparse variant, in both the $p < n$ and $p > n$ regimes. For instance, with 20\% vertical outliers and $p = 100$ ($p_{\text{signal}} = 20$, $p_{\text{noise}} = 80$), the MSEP of sparse RTB is 1.63, compared to 4.85 for sparse twoblock.

\begin{figure}[ht]
\centering
\includegraphics[width=\textwidth]{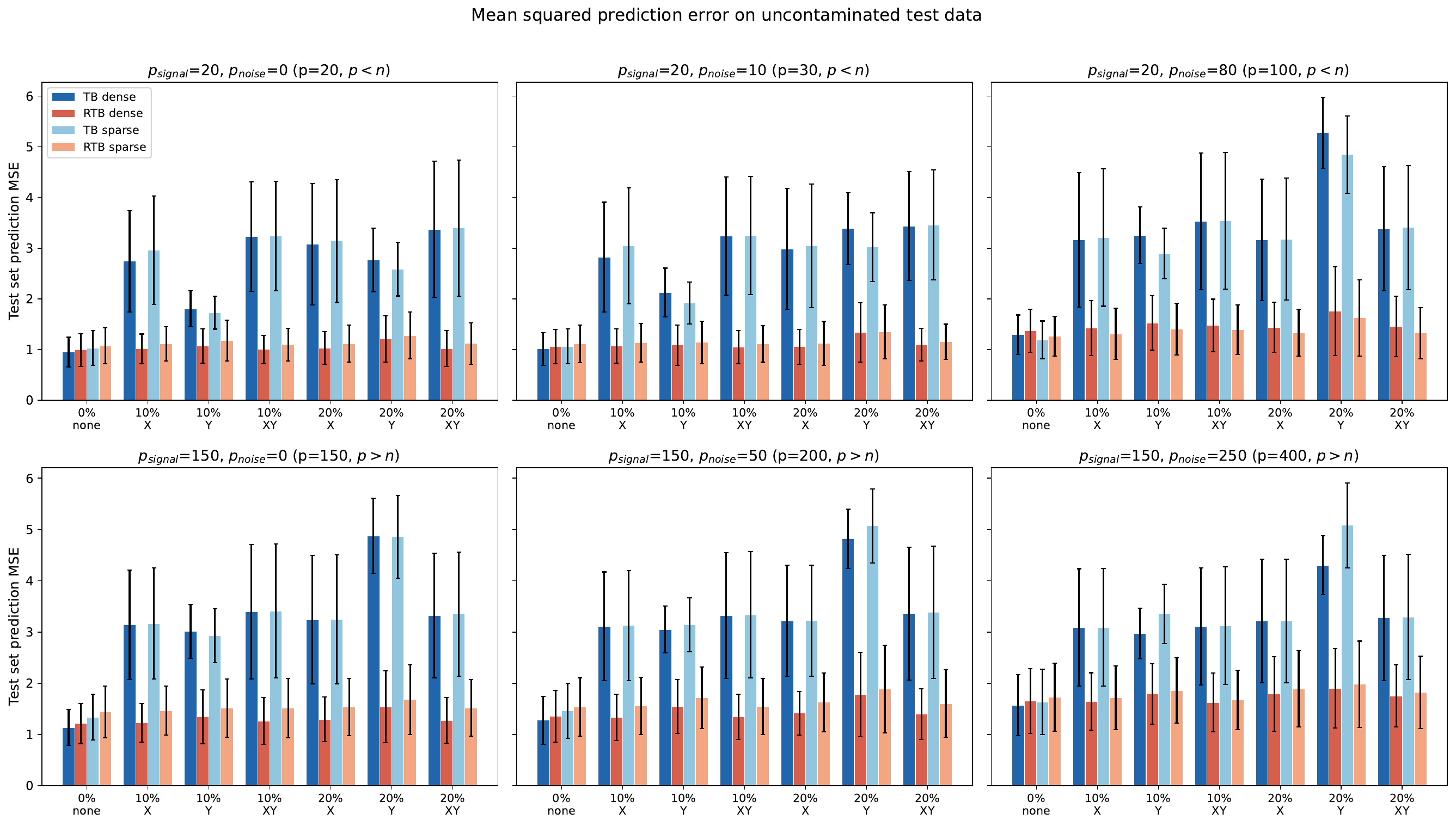}
\caption{Mean squared prediction error on uncontaminated test sets ($n_{\text{test}} = 1000$) across 42 simulation scenarios (200 repeats each). Colours as in Figure~\ref{fig:sim_mse}. Top row: $p \leq n$; bottom row: $p > n$. Error bars indicate one standard deviation.}
\label{fig:sim_pred}
\end{figure}
}

\begin{figure}[ht]
\centering
\includegraphics[width=\textwidth]{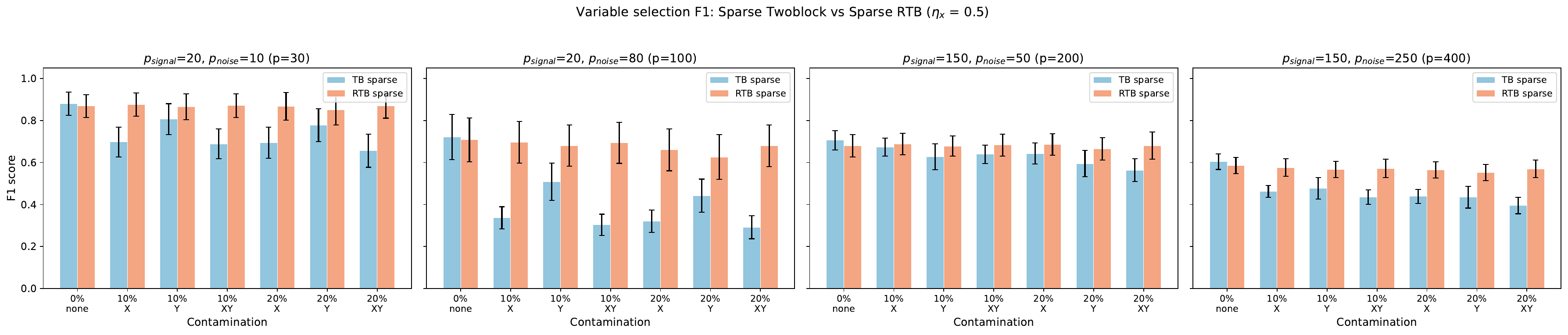}
\caption{Variable selection $F_1$ score for sparse twoblock (light blue) and sparse RTB (salmon) across contamination scenarios, for configurations with noise variables ($\eta_x = 0.5$). Higher is better.}
\label{fig:sim_f1}
\end{figure}

\paragraph{Variable selection accuracy.}
For the sparse methods, Figure~\ref{fig:sim_f1} reports the $F_1$ score for variable selection---the harmonic mean of precision and recall for identifying the true signal variables among a mixture of signal and noise variables.

\replaced{On clean data, both sparse methods achieve nearly identical $F_1$ scores (e.g.\ 0.72 for sparse twoblock and 0.71 for sparse RTB with $p_{\text{signal}} = 20$, $p_{\text{noise}} = 80$), confirming that robust preprocessing does not impair variable selection in the absence of outliers.
Under contamination, a striking pattern emerges: the $F_1$ of sparse twoblock degrades substantially for all types of contamination, most strongly when $\mathbf{X}$ outliers or joint $\mathbf{X}$--$\mathbf{Y}$ outliers are present (e.g.\ from 0.72 to 0.29--0.34 for $p_{\text{signal}} = 20$, $p_{\text{noise}} = 80$, and from 0.60 to 0.40--0.46 for $p_{\text{signal}} = 150$, $p_{\text{noise}} = 250$), whereas sparse RTB maintains its $F_1$ close to the clean-data level in all scenarios (0.63--0.70 and 0.55--0.58, respectively).
This demonstrates that RTB not only protects coefficient estimation but also preserves the variable selection performance of the sparse method under contamination. Vertical outliers degrade the variable selection of sparse twoblock as well, albeit to a lesser extent (e.g.\ from 0.72 to 0.44--0.51 for $p_{\text{signal}} = 20$, $p_{\text{noise}} = 80$): although sparsity is applied to the $\mathbf{X}$-block weights, these weights are computed from the covariance with the contaminated $\mathbf{Y}$ block.}{On clean data, both sparse methods achieve identical $F_1$ scores, confirming that robust preprocessing does not impair variable selection in the absence of outliers.
Under contamination, however, a striking pattern emerges: sparse twoblock's $F_1$ degrades substantially when $\mathbf{X}$ outliers or joint $\mathbf{X}$---$\mathbf{Y}$ outliers are present (e.g.\ from 0.85 to 0.57--0.67 for $p_{\text{signal}} = 150$, $p_{\text{noise}} = 50$ at 10--20\% contamination), while sparse RTB maintains its $F_1$ near the clean-data level ($\approx 0.85$).
This demonstrates that RTB not only protects coefficient estimation but also preserves the variable selection consistency of the sparse method under contamination.
Interestingly, $\mathbf{Y}$-only outliers do not degrade variable selection for either method, since the sparsity is applied to the $\mathbf{X}$-block weights.}

\paragraph{\added[R1.15]{Fewer responses than latent factors.}}
\added[R1.15]{In the design above, $q > k$, such that the rule $h_y = \min(k, q)$ always yields $h_y = k$. To assess the regime in which the response block has fewer variables than the dimension of the latent structure, the complete simulation (all 42 scenarios, 200 repeats each) was repeated with $q = 2$ responses and $k = 3$. Here, the systematic part $\mathbf{Z}\mathbf{C}$ of $\mathbf{Y}$ spans all $q = 2$ dimensions, so the $\mathbf{Y}$ block cannot be reduced, and the rule gives $h_y = q = 2$. All methods were fit both with $h_y = 2$ and with one $\mathbf{Y}$ component too few ($h_y = 1$). Table~\ref{tab:sim_qk} shows the prediction errors for two of the six dimensionality settings. The comparison between RTB and twoblock is qualitatively the same as for $q = 4$. On clean data, the MSEP ratio of RTB to twoblock is 1.06--1.10 (dense) and 1.06--1.12 (sparse); under contamination, it ranges from 0.19 to 0.56 (dense) and from 0.24 to 0.56 (sparse), across all six dimensionality settings. The MSE of the coefficients and the $F_1$ scores (not shown) follow the same pattern as for $q = 4$. Underestimating $h_y$ is costly: on clean data, $h_y = 1$ increases the MSEP by 26--78\% compared to $h_y = 2$, for all four methods, and for RTB this holds under contamination as well (20--73\%). Only for classical twoblock under vertical outliers does $h_y = 1$ yield a lower prediction error, since discarding a $\mathbf{Y}$ direction then also discards part of the outlying variation; this partial protection remains far behind that of RTB. The rule $h_y = \min(k, q)$ is therefore appropriate in this regime as well, and the robustness of RTB does not depend on whetther $q$ exceeds $k$.}

\begin{table}[ht]
\centering
\caption{\added[R1.15]{Simulation with fewer responses than latent factors ($q = 2 < k = 3$): mean squared prediction error on uncontaminated test sets (200 repeats), for dense twoblock (TB) and dense RTB with the rule $h_y = \min(k, q) = 2$ and with one $\mathbf{Y}$ component too few ($h_y = 1$); $h_x = k = 3$ throughout.}}
\label{tab:sim_qk}
\small
\begin{tabular}{l|cccc|cccc}
\toprule
 & \multicolumn{4}{c|}{\added[R1.15]{$p_{\text{signal}} = 20$, $p_{\text{noise}} = 80$}} & \multicolumn{4}{c}{\added[R1.15]{$p_{\text{signal}} = 150$, $p_{\text{noise}} = 250$}} \\
 & \multicolumn{2}{c}{\added[R1.15]{$h_y = 2$}} & \multicolumn{2}{c|}{\added[R1.15]{$h_y = 1$}} & \multicolumn{2}{c}{\added[R1.15]{$h_y = 2$}} & \multicolumn{2}{c}{\added[R1.15]{$h_y = 1$}} \\
\added[R1.15]{Contamination} & \added[R1.15]{TB} & \added[R1.15]{RTB} & \added[R1.15]{TB} & \added[R1.15]{RTB} & \added[R1.15]{TB} & \added[R1.15]{RTB} & \added[R1.15]{TB} & \added[R1.15]{RTB} \\
\midrule
\added[R1.15]{Clean} & \added[R1.15]{1.33} & \added[R1.15]{1.46} & \added[R1.15]{1.85} & \added[R1.15]{2.14} & \added[R1.15]{1.60} & \added[R1.15]{1.74} & \added[R1.15]{2.01} & \added[R1.15]{2.34} \\
\added[R1.15]{10\% $\mathbf{X}$} & \added[R1.15]{3.01} & \added[R1.15]{1.50} & \added[R1.15]{3.06} & \added[R1.15]{2.14} & \added[R1.15]{3.17} & \added[R1.15]{1.75} & \added[R1.15]{3.18} & \added[R1.15]{2.32} \\
\added[R1.15]{10\% $\mathbf{Y}$} & \added[R1.15]{5.27} & \added[R1.15]{1.69} & \added[R1.15]{4.36} & \added[R1.15]{2.25} & \added[R1.15]{3.71} & \added[R1.15]{1.71} & \added[R1.15]{3.40} & \added[R1.15]{2.18} \\
\added[R1.15]{10\% $\mathbf{X}\mathbf{Y}$} & \added[R1.15]{3.00} & \added[R1.15]{1.34} & \added[R1.15]{3.07} & \added[R1.15]{1.82} & \added[R1.15]{3.27} & \added[R1.15]{1.72} & \added[R1.15]{3.32} & \added[R1.15]{2.23} \\
\added[R1.15]{20\% $\mathbf{X}$} & \added[R1.15]{3.30} & \added[R1.15]{1.61} & \added[R1.15]{3.31} & \added[R1.15]{2.19} & \added[R1.15]{3.12} & \added[R1.15]{1.76} & \added[R1.15]{3.12} & \added[R1.15]{2.33} \\
\added[R1.15]{20\% $\mathbf{Y}$} & \added[R1.15]{9.14} & \added[R1.15]{1.73} & \added[R1.15]{6.73} & \added[R1.15]{2.30} & \added[R1.15]{6.04} & \added[R1.15]{1.92} & \added[R1.15]{4.95} & \added[R1.15]{2.31} \\
\added[R1.15]{20\% $\mathbf{X}\mathbf{Y}$} & \added[R1.15]{3.25} & \added[R1.15]{1.45} & \added[R1.15]{3.35} & \added[R1.15]{2.03} & \added[R1.15]{3.40} & \added[R1.15]{1.82} & \added[R1.15]{3.48} & \added[R1.15]{2.35} \\
\bottomrule
\end{tabular}
\end{table}

\section{Applications to real world data}\label{sec:examples}

In this section, the utility of the new methods will be illustrated in a set of two examples. The first data set contains a moderate number of cases, yet has more variables then cases. This data set is included, because it was already identified in the seminal paper on twoblock dimension reduction \citep{cook2023} as a data set that illustrates the benefits of simultaneous dimension reduction well. Moreover, \cite{serneels2025} showed that sparse twoblock dimension's intrinsic variable selection improves upon the results in \cite{cook2023}. Moreover, the data set is known to contain outliers, which were removed in both earlier publications. This section will show that robust sparse twoblock does approximately as well as the two classical methods in spite of the presence of outliers and for one of the four response variables, even outperforms the previously published results.   

The second data set considers a set of sensor readings from gas turbine operations. It has far more cases than variables. Variables were selected to be representative, yet it is plausible that one or two variables contribute little to the modeling target. The data set's response variables are known to generate natural vertical outliers, which is conducive to an illustrative comparison between the methods proposed herein. 

\subsection{Cookie dough NIR data}\label{sec:cookie}

The cookie dough data set \citep{osborne1984} consists of near-infrared (NIR) spectra ($p = 700$ wavelengths) measured on dough pieces, with laboratory measurements of four analytes (Fat, Sucrose, Flour, Water) as the response.
The training set contains 40 samples and the test set 32 samples.
This data set is known to contain atypical observations: case~23 in the training set and case~21 in the test set have been identified as outliers.

For the non-robust methods, these cases are removed and the fit is reported for the clean data ($n_{\text{train}} = 39$, $n_{\text{test}} = 31$).
For RTB, all cases are retained for training, so as to allow the reweighting to handle the outliers automatically.

\paragraph{Methods compared.}
We compare PLS2, PLS1 (per-response), twoblock (dense and sparse), and RTB (dense and sparse).
\added[R1.7, R1.8]{Moreover, the closest existing alternatives discussed in Section~\ref{sec:related} are included as benchmarks: sparse PLS2 (SPLS2, \citealp{chun2010}), CCA, sparse CCA by penalized matrix decomposition \citep{witten2009}, and SMPCovR \citep{park2024}. CCA and sparse CCA predict $\mathbf{Y}$ by least squares regression on the $\mathbf{X}$ canonical variates. SMPCovR is tuned by the procedure proposed by its authors (5-fold cross-validation over the weighting parameter $\alpha$ and both lasso penalties, for each number of components $R$).}
For the classical methods, hyperparameters are selected by cross-validation on the clean data.
For RTB, hyperparameters are selected by cross-validation on the full (contaminated) data, with Hampel weighting (cutoffs $p_1 = 0.75$, $p_2 = 0.90$, $p_3 = 0.95$) and median centring.
\added[R1.18]{In all cases, cross-validation is a joint grid search: every combination of the hyperparameters in the grid (the numbers of components, the sparsity parameters where applicable, and the scaling option) is evaluated, and the combination with the best cross-validated $R^2$ is refitted on the complete training set. No sequential tuning is involved. For sparse twoblock, the configuration was obtained earlier by such a joint grid search over a finer grid.}
\added[R2.14]{The scaling used for RTB divides each variable, after centring by the median, by its k-step LTS location estimate rather than by a scale estimate. This amounts to a robust version of level scaling \citep{vandenberg2006}: each wavelength is expressed relative to its typical absorbance level, and each analyte relative to its typical content.}

Cross-validation selected the following configurations:
\begin{itemize}
    \item PLS2: 6 components, no scaling.
    \item PLS1: 7 (Fat), 6, 6, 7 (Water) components.
    \item Twoblock dense: $h_x = 6$, $h_y = 2$, no scaling.
    \item Twoblock sparse: $h_x = 9$, $h_y = 2$, $\eta_x = 0.5$, standard scaling.
    \item \added[R1.7]{SPLS2: 6 components, $\eta = 0.3$, no scaling (693 wavelengths retained).}
    \item \added[R1.7]{CCA: 1 component, standard scaling.}
    \item \added[R1.7]{Sparse CCA: 4 components, $\ell_1$ bounds $0.5\sqrt{p}$ and $0.6\sqrt{q}$, no scaling (419 wavelengths retained).}
    \item \added[R1.8]{SMPCovR: $R = 10$, $\alpha = 0.1$, standard scaling (573 wavelengths and all four responses retained).}
    \item RTB dense: $h_x = 11$, $h_y = 2$, \replaced[R2.14]{robust level scaling}{kstepLTS scaling}.
    \item RTB sparse: $h_x = 8$, $h_y = 2$, $\eta_x = 0.3$, \replaced[R2.14]{robust level scaling}{kstepLTS scaling}.
\end{itemize}
\added[R2.14]{The number of $\mathbf{X}$ components selected for dense RTB is higher than for twoblock and PLS2. With only 40 training cases, however, the selected number of components depends considerably on how the cases are allocated to the folds. Repeating the cross-validation of dense RTB for 50 random 5-fold splits, the median selected value is $h_x = 7$, and $h_x = 11$ was selected for only one split. Moreover, the different scaling implies that the components are extracted from a different covariance structure, such that the numbers of components are not directly comparable between the methods. Over these 50 splits, the test set $R^2$ of dense RTB (mean over the four analytes) ranges from $0.81$ to $0.93$, and cases~23 and~24 are downweighted for every split.}

\paragraph{Results.}

Table~\ref{tab:cookie} reports test set $R^2$ for each method and response variable.
The non-robust methods are trained and evaluated on the clean data (39 training, 31 test samples), while RTB is trained on the full data (40 training samples, including the outlier) and evaluated on the clean test set (31 samples).

\begin{table}[ht]
\centering
\caption{Cookie dough data: test set $R^2$ by method and response variable.
Non-robust methods trained on clean data; RTB trained on full data including outliers.}
\label{tab:cookie}
\begin{tabular}{lcccc|c}
\toprule
Method & Fat & Sucrose & Flour & Water & Mean \\
\midrule
PLS2 (clean)            & 0.550 & 0.948 & 0.746 & 0.658 & 0.725 \\
PLS1 (clean)            & 0.979 & 0.935 & 0.730 & 0.916 & 0.890 \\
Twoblock dense (clean)  & 0.596 & 0.940 & 0.821 & 0.379 & 0.684 \\
Twoblock sparse (clean) & 0.930 & \replaced[R1.7]{0.962}{\textbf{0.962}} & \textbf{0.931} & \replaced[R1.7]{0.948}{\textbf{0.948}} & \textbf{0.943} \\
\added[R1.7]{SPLS2 (clean)}       & \added[R1.7]{0.555} & \added[R1.7]{0.949} & \added[R1.7]{0.750} & \added[R1.7]{0.662} & \added[R1.7]{0.729} \\
\added[R1.7]{CCA (clean)}         & \added[R1.7]{0.585} & \added[R1.7]{0.014} & \added[R1.7]{$-0.034$} & \added[R1.7]{0.178} & \added[R1.7]{0.186} \\
\added[R1.7]{Sparse CCA (clean)}  & \added[R1.7]{0.473} & \added[R1.7]{\textbf{0.966}} & \added[R1.7]{0.908} & \added[R1.7]{0.511} & \added[R1.7]{0.715} \\
\added[R1.8]{SMPCovR (clean)}     & \added[R1.8]{0.951} & \added[R1.8]{0.679} & \added[R1.8]{0.316} & \added[R1.8]{\textbf{0.961}} & \added[R1.8]{0.727} \\
\midrule
RTB dense (full)        & 0.933 & 0.908 & 0.839 & 0.823 & 0.876 \\
RTB sparse (full)       & \textbf{0.980} & 0.914 & 0.876 & 0.904 & 0.918 \\
\bottomrule
\end{tabular}
\end{table}

Several observations stand out.
At first, the dense twoblock on clean data performs poorly for Fat and Water ($R^2 = 0.596$ and $0.379$), suggesting that even after removing the known outliers, the classical method is sensitive to the remaining data structure. Secondly, the sparse twoblock on clean data achieves the best overall performance (mean $R^2 = 0.943$), demonstrating the value of variable selection. \added[R1.7, R1.8]{None of the additional benchmarks comes close to this on all four analytes. SPLS2 performs on a par with PLS2 (mean $R^2 = 0.729$), as its cross-validated sparsity retains 693 of the 700 wavelengths. CCA breaks down since $p > n$, such that the canonical correlations are trivially equal to one. Sparse CCA and SMPCovR attain the best $R^2$ for Sucrose and Water, respectively, but perform poorly on the remaining analytes (mean $R^2$ of $0.715$ and $0.727$).} Thirdly, RTB dense, trained on the \emph{full} data including the outlier, achieves a mean test $R^2$ of $0.876$---substantially better than the dense twoblock on clean data.
Finally, RTB sparse on full data achieves strong performance across all four analytes (mean $R^2 = 0.918$), competitive with the sparse twoblock on clean data, without requiring prior outlier identification or removal. This at once illustrates the utility of the method, as well as its outlier robustness and high efficiency, reaching results close to classical estimators for uncontaminated data. 

\paragraph{Case weight diagnostics.}

\begin{figure}[ht]
\centering
\includegraphics[width=0.9\textwidth]{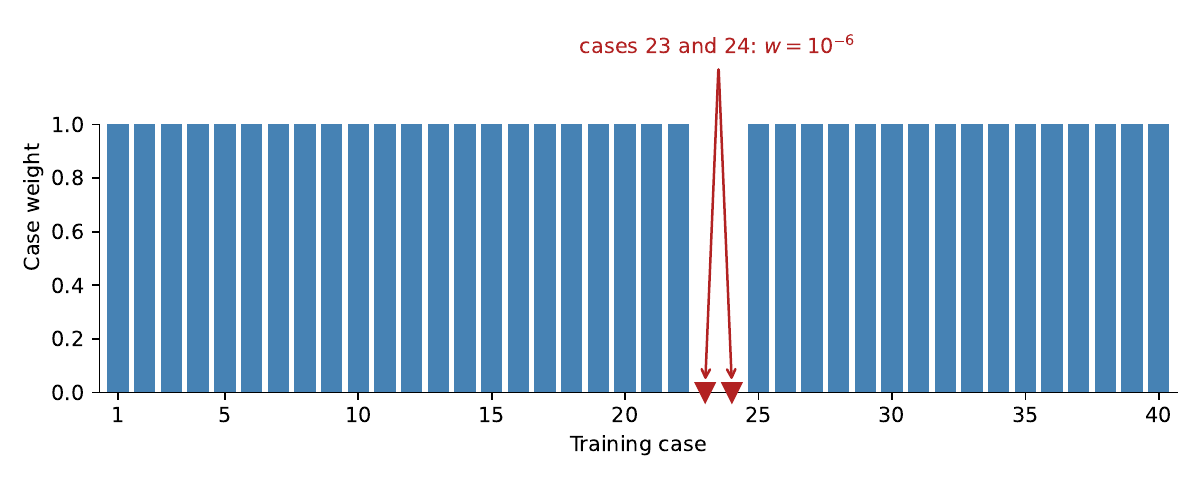}
\caption{RTB dense case weights for the cookie dough training data. \replaced[R1.B, R2.15]{Cases~23 and~24 (marked by arrows) are assigned a weight of $10^{-6}$, such that no bar is visible for them; all other cases receive full weight.}{Case~23 (red) and case~24 are assigned near-zero weights.}}
\label{fig:caseweights_dense}
\end{figure}

Figure~\ref{fig:caseweights_dense} shows the combined case weights from the dense RTB model.
Case~23 receives a weight of $10^{-6}$ (effectively zero), confirming that RTB automatically identifies and downweights this known outlier.
Case~24 is also downweighted\added[R1.B, R2.15]{ to the same weight}, suggesting it may also be atypical. All other cases receive full weight ($w = 1$).

\begin{figure}[ht]
\centering
\includegraphics[width=0.9\textwidth]{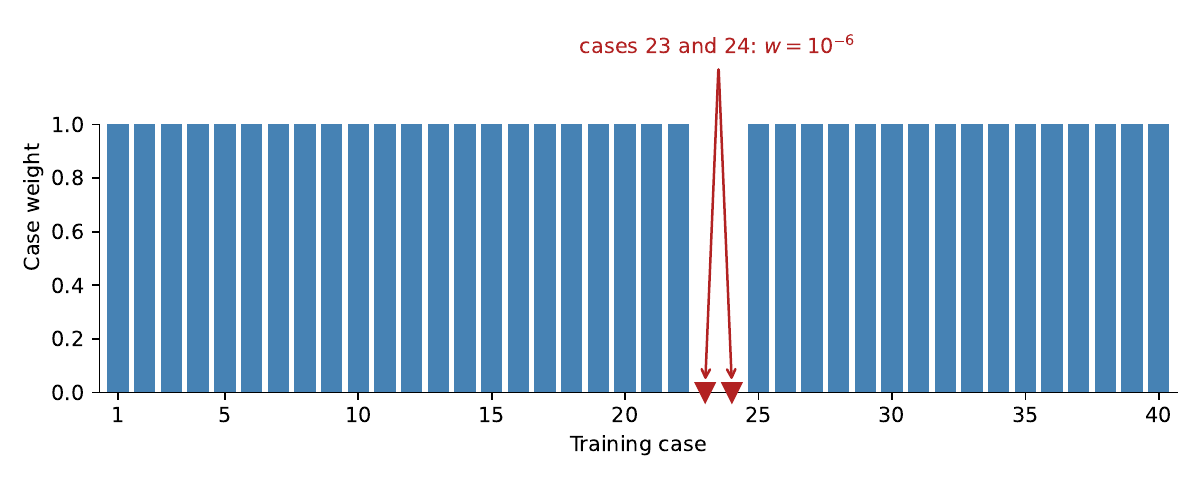}
\caption{RTB sparse case weights for the cookie dough training data. Cases~23 and~24 \replaced[R1.B, R2.15]{(marked by arrows) are again assigned a weight of $10^{-6}$}{(red: case~23) are again assigned near-zero weights}.}
\label{fig:caseweights_sparse}
\end{figure}

The sparse RTB model produces the same weight pattern (Figure~\ref{fig:caseweights_sparse}), confirming that the outlier identification is consistent across the dense and sparse variants. The fact that the sparse robust method performs better than the classical methods for the fat variable may indicate that case 24 could be a vertical outlier for that variable only. However, there is no way to practically verify that assumption.    

\paragraph{Variable selection.}

\begin{figure}[ht]
\centering
\includegraphics[width=0.9\textwidth]{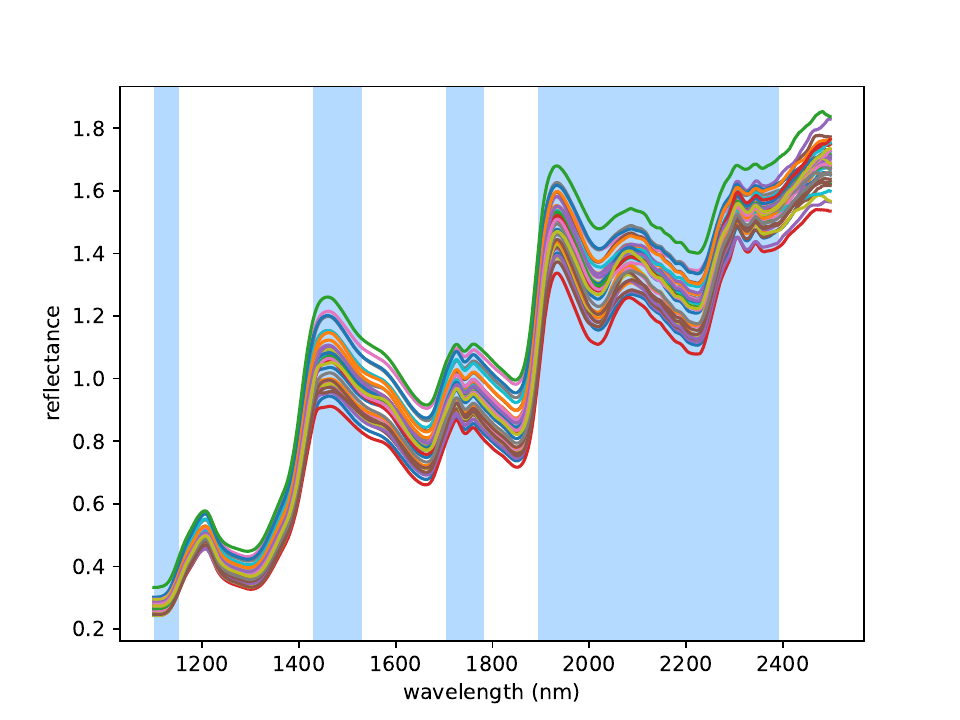}
\caption{Cookie dough NIR spectra with eliminated wavelength regions (blue shading) from the sparse RTB model.}
\label{fig:eliminated}
\end{figure}

The sparse RTB model ($\eta_x = 0.3$, $h_x = 8$) eliminates 368 of 700 wavelengths (52.6\%), retaining 332 wavelengths (47.4\%).
Figure~\ref{fig:eliminated} shows the NIR spectra overlaid with the eliminated wavelength regions (shaded in blue).
Four main regions are eliminated: the initial baseline region (wavelengths 0--26), and three regions in the mid- to high-wavelength range (165--215, 302--341, 397--646).
These correspond to spectral regions with low signal-to-noise ratio or limited discriminative power for the four analytes. It is noted that this variable selection performance is virtually indistinguishable from the one obtained from non-robust sparse twoblock dimension reduction, first reported in \citep{serneels2025}.

\subsection{Gas turbine data}\label{sec:gas}

The gas turbine CO and NOx emission data set \citep{kaya2019} contains 36{,}733 hourly sensor readings from a gas turbine in a combined cycle power plant, collected between 2011 and 2015. The predictor block consists of $p = 9$ process variables: ambient temperature (AT), ambient pressure (AP), ambient humidity (AH), air filter differential pressure (AFDP), gas turbine exhaust pressure (GTEP), turbine inlet temperature (TIT), turbine after temperature (TAT), turbine energy yield (TEY), and compressor discharge pressure (CDP).
The bivariate response consists of CO and NOx emission concentrations, which are known to generate natural vertical outliers.

\begin{figure}[ht]
\centering
\includegraphics[width=0.7\textwidth]{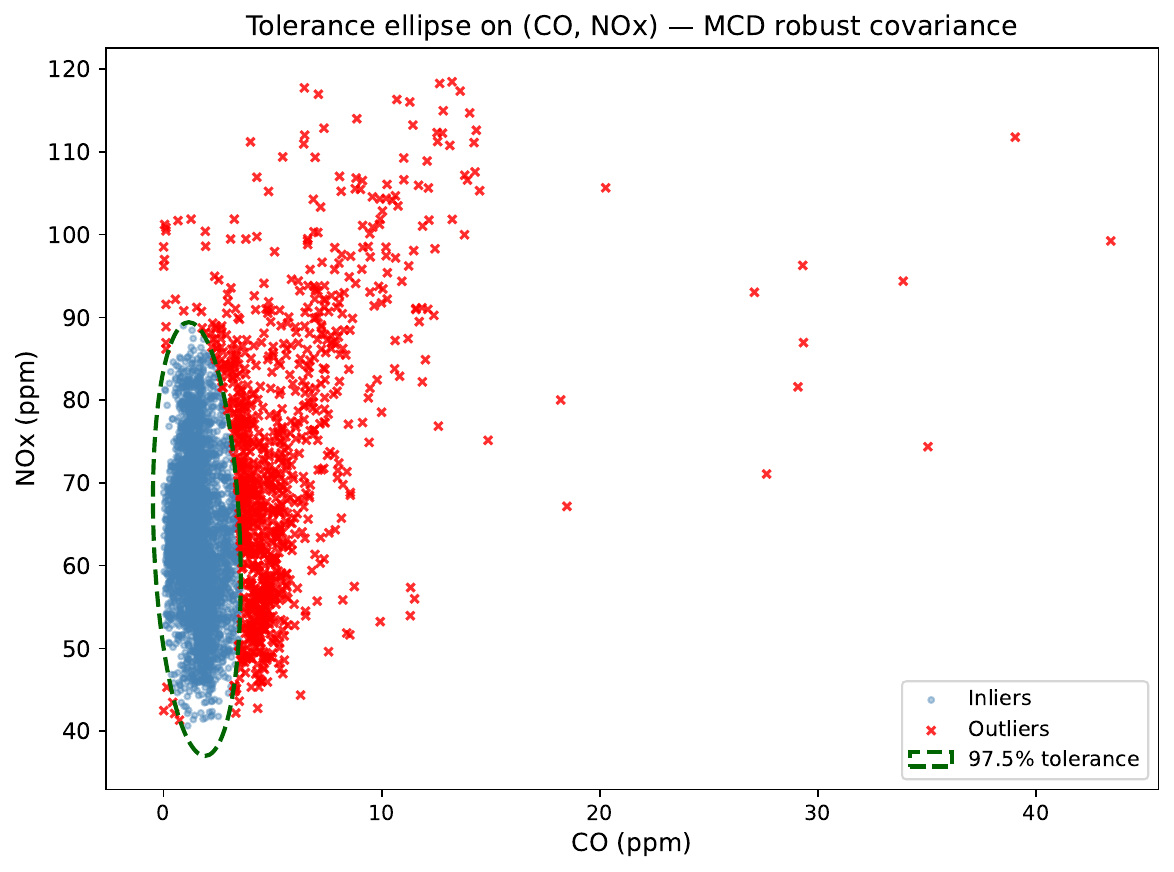}
\caption{Bivariate (CO, NOx) scatter plot with 97.5\% MCD tolerance ellipse. Red crosses: observations classified as outliers; blue: inliers used for the clean training and test sets.}
\label{fig:gas_ellipse}
\end{figure}

To identify these natural outliers, a 97.5\% tolerance ellipse based on the Minimum Covariance Determinant (MCD) estimator \citep{rousseeuw1985} was fitted to the bivariate response space.
Out of a random subsample of 5{,}000 observations, this identified 1{,}064 outliers (21.3\%), which were separated from 3{,}936 inliers.
The inliers were split 70/30 into a clean training set ($n_{\text{train}} = 2{,}755$) and a clean test set ($n_{\text{test}} = 1{,}181$).
A contaminated training set was then constructed by injecting 306 of the detected outliers (targeting $\approx$10\% contamination), with care taken to include cases that are extreme in both CO and NOx individually.
The tolerance ellipse and the resulting outlier classification are shown in Figure~\ref{fig:gas_ellipse}.

\paragraph{Methods compared.}
\replaced[R1.7, R1.8]{Nine}{Five} methods were compared, each with hyperparameters selected by 5-fold cross-validation on the respective training set:
\begin{itemize}
    \item PLS2: 9 components.
    \item Twoblock dense: $h_x = 9$, $h_y = 2$, no scaling (clean); $h_x = 7$, $h_y = 2$ (contaminated).
    \item Twoblock sparse: $h_x = 9$, $h_y = 2$, $\eta_x = \replaced{0.7}{0.3}$ (clean); $h_x = 5$, $h_y = 2$, $\eta_x = 0.7$ (contaminated).
    \item \added[R1.7]{SPLS2: 9 components, $\eta = 0.1$, all 9 predictors retained (clean); 3 components, $\eta = 0.7$, 4 predictors retained (contaminated).}
    \item \added[R1.7]{CCA: 2 components.}
    \item \added[R1.7]{Sparse CCA: 2 components, $\ell_1$ bounds $0.35\sqrt{p}$ and $0.75\sqrt{q}$, 4 predictors retained (both training sets).}
    \item \added[R1.8]{SMPCovR: $R = 4$ (clean) and $R = 6$ (contaminated), $\alpha = 0.9$, all predictors and both responses retained.}
    \item RTB dense: $h_x = 9$, $h_y = 2$ (clean); $h_x = 7$, $h_y = 1$ (contaminated).
    \item RTB sparse: $h_x = 9$, $h_y = 2$, $\eta_x = \replaced{0.7}{0.3}$ (clean); $h_x = 5$, $h_y = 2$, $\eta_x = 0.7$ (contaminated).
\end{itemize}

\paragraph{Results.}

Table~\ref{tab:gas} reports the test set MSE for each method under both training scenarios. Since the CO and NOx variables differ substantially in variance ($\sigma^2_{\text{CO}} \approx 5.7$, $\sigma^2_{\text{NOx}} \approx 135$), a weighted mean MSE is also reported using inverse-variance weights, ensuring both responses contribute proportionally to the overall score.

\begin{table}[ht]
\centering
\caption{Gas turbine data: test set MSE by method and response variable.
The weighted mean uses inverse-variance weights (CO: 0.960, NOx: 0.040).}
\label{tab:gas}
\begin{tabular}{l|ccc|ccc}
\toprule
 & \multicolumn{3}{c|}{Clean training} & \multicolumn{3}{c}{Contaminated training} \\
Method & CO & NOx & Wt.\ Mean & CO & NOx & Wt.\ Mean \\
\midrule
PLS2       & 0.318 & 30.40 & 1.532 & 1.111 & 47.26 & 2.974 \\
TB dense   & 0.318 & 30.40 & 1.532 & 1.121 & 47.56 & 2.997 \\
TB sparse  & 0.318 & 30.40 & 1.532 & 1.114 & 60.43 & 3.509 \\
\added[R1.7]{SPLS2}      & \added[R1.7]{0.318} & \added[R1.7]{30.40} & \added[R1.7]{1.532} & \added[R1.7]{1.151} & \added[R1.7]{70.51} & \added[R1.7]{3.951} \\
\added[R1.7]{CCA}        & \added[R1.7]{0.344} & \added[R1.7]{53.12} & \added[R1.7]{2.475} & \added[R1.7]{0.672} & \added[R1.7]{64.18} & \added[R1.7]{3.236} \\
\added[R1.7]{Sparse CCA} & \added[R1.7]{0.491} & \added[R1.7]{56.28} & \added[R1.7]{2.743} & \added[R1.7]{1.138} & \added[R1.7]{68.51} & \added[R1.7]{3.858} \\
\added[R1.8]{SMPCovR}    & \added[R1.8]{0.318} & \added[R1.8]{30.62} & \added[R1.8]{1.541} & \added[R1.8]{1.119} & \added[R1.8]{46.41} & \added[R1.8]{2.948} \\
RTB dense  & \textbf{0.318} & \textbf{30.38} & \textbf{1.532} & 0.566 & \textbf{34.99} & 1.956 \\
RTB sparse & \textbf{0.318} & \textbf{30.38} & \textbf{1.532} & \textbf{0.349} & 38.97 & \textbf{1.908} \\
\bottomrule
\end{tabular}
\end{table}

On clean data, \replaced{PLS2, both twoblock variants, SPLS2 and SMPCovR achieve virtually identical MSE, and RTB incurs no measurable efficiency loss.}{all methods except PRM achieve virtually identical MSE, with RTB incurring no measurable efficiency loss. It is still noted that the efficiency loss for PRM is limited.}
\added[R1.7, R1.8]{CCA and sparse CCA are clearly less accurate than the other methods already on clean data, in particular for NOx.}
Under contamination, the non-robust methods (PLS2, TB dense, TB sparse\added[R1.7, R1.8]{, SPLS2, sparse CCA, SMPCovR}) suffer a roughly twofold MSE increase for CO, while the RTB variants remain close to their clean-data performance. Notably, RTB sparse on contaminated data achieves a CO MSE of 0.349, very close to the result for clean data (0.318), while also providing the best result for the weighted mean MSE for both variables, indicating that sparse RTB yields the best simultaneous multivariate model for both emissions in the presence of outliers.

\paragraph{\added[R2.16]{Model complexity under contamination.}}
\added[R2.16]{On the contaminated training set, cross-validation selects fewer $\mathbf{X}$ components ($h_x = 7$ or $5$) than on the clean training set ($h_x = 9 = p$), for the classical as well as for the robust methods. This is not a property of the data, but of the cross-validation criterion: the mean squared prediction error is computed on all validation cases, including the vertical outliers. Since no model can (or should) predict these, their large residuals dominate the criterion, which masks the comparatively small gain from the higher order components. Indeed, when the cross-validated prediction error is computed on the inlying validation cases only, it decreases monotonically in $h_x$ for all four twoblock and RTB variants, and $h_x = 9$ is selected, as on the clean data. The test set MSE likewise is lowest at $h_x = 9$ for all four variants. For sparse RTB, $h_x = 9$ would reduce the weighted mean test MSE from $1.908$ to $1.599$, and for dense RTB from $1.956$ to $1.924$. The results reported in Table~\ref{tab:gas} are thus conservative for RTB. It is concluded that, while RTB estimates the model robustly, its hyperparameters should be selected by a robust cross-validation criterion as well when the data are contaminated.}

\paragraph{Case weight diagnostics.}

\begin{figure}[ht]
\centering
\includegraphics[width=0.9\textwidth]{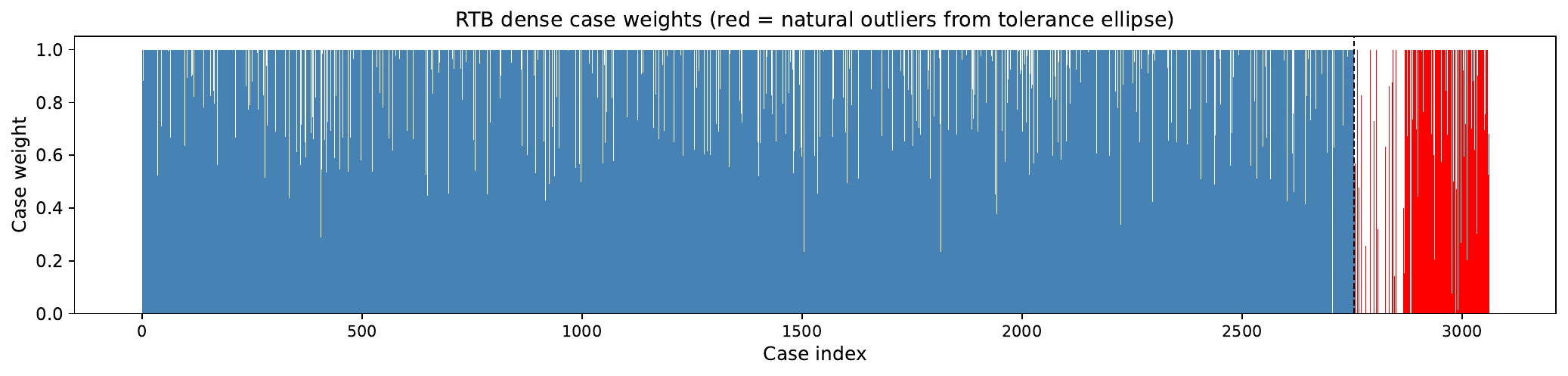}
\caption{RTB dense case weights for the gas turbine contaminated training set. Red bars indicate natural outliers identified by the MCD tolerance ellipse.}
\label{fig:gas_caseweights}
\end{figure}

Figure~\ref{fig:gas_caseweights} shows the combined case weights from the RTB dense model fitted on the contaminated training set. The 306 injected outliers (red bars, appended after the 2{,}755 clean cases) receive substantially reduced weights, with 47\% fully downweighted ($w < 0.5$).
The remaining outliers receive partial weight, consistent with the fact that not all observations outside the tolerance ellipse are equally extreme.

\begin{figure}[ht]
\centering
\includegraphics[width=0.9\textwidth]{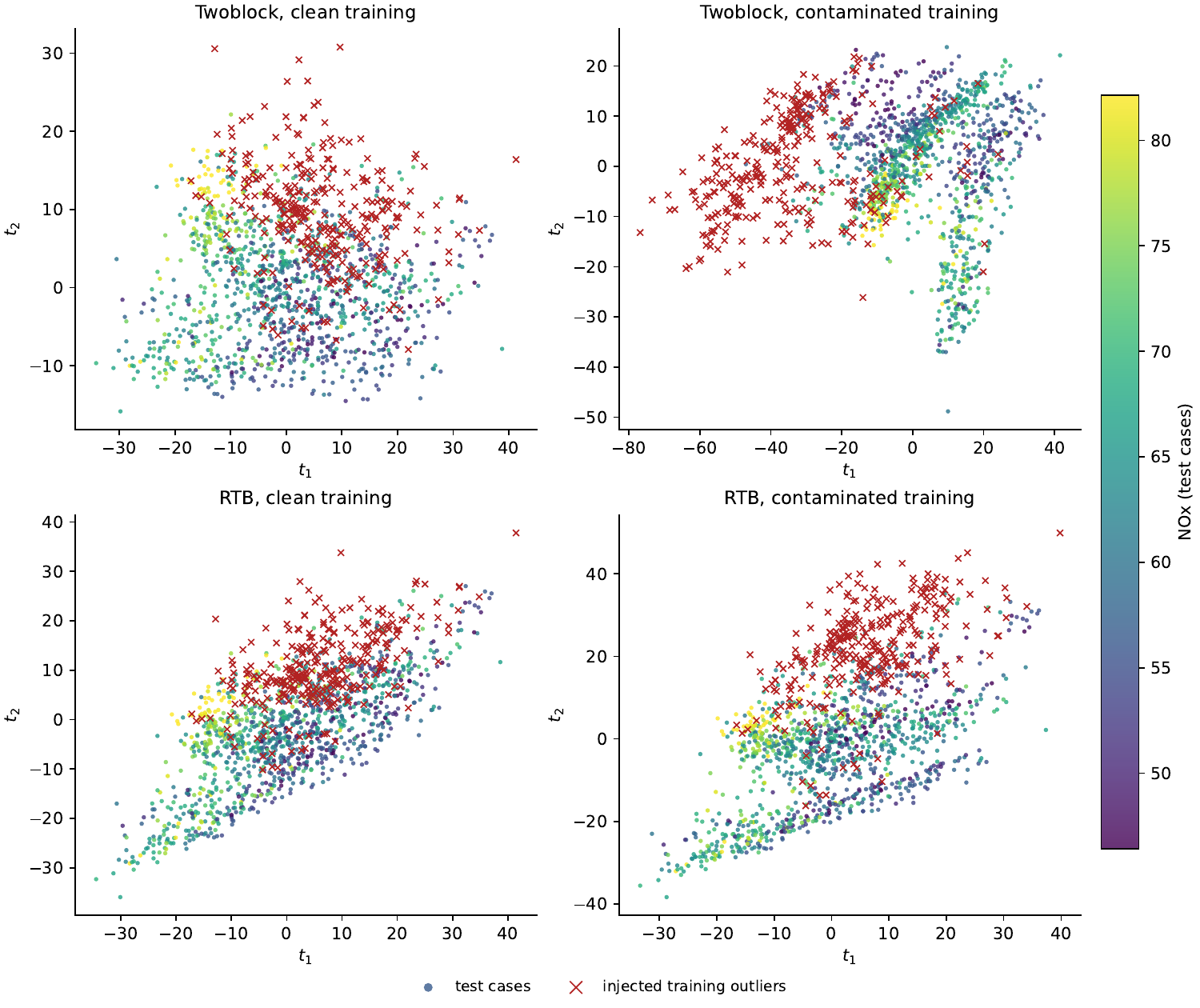}
\caption{\added[R2.17]{Gas turbine data: projection of the clean test set (coloured by NOx) and of the 306 injected training outliers (red crosses) onto the first two $\mathbf{X}$ components of dense twoblock (top) and dense RTB (bottom), trained on the clean (left) and on the contaminated (right) training set, with $h_x = 7$, $h_y = 2$ and no scaling throughout. Axis signs are aligned with those of the same method trained on clean data.}}
\label{fig:gas_scores}
\end{figure}

\paragraph{\added[R2.17]{Projection of the test set.}}
\added[R2.17]{Figure~\ref{fig:gas_scores} shows the projection of the clean test set onto the first two $\mathbf{X}$ components of dense twoblock and dense RTB, each trained on the clean and on the contaminated training set. To isolate the effect of the contamination, a common configuration is used for all four models ($h_x = 7$, $h_y = 2$, no scaling), and the 306 injected training outliers are projected as well. On clean data, both methods yield a similar picture. Trained on the contaminated data, the twoblock score space is distorted: its first component is drawn towards the outliers, which are spread out along it, while the test cases are compressed. The correlation between the test set scores of twoblock trained on contaminated and on clean data is only $0.19$ for the first and $0.02$ for the second component. In contrast, the RTB projection of the test set is virtually unaffected by the contamination (correlations of $0.999$ and $0.954$), and the outliers are separated more clearly from the bulk of the data than in the RTB model trained on clean data.}

\paragraph{Variable selection.}

\begin{figure}[ht]
\centering
\includegraphics[width=0.9\textwidth]{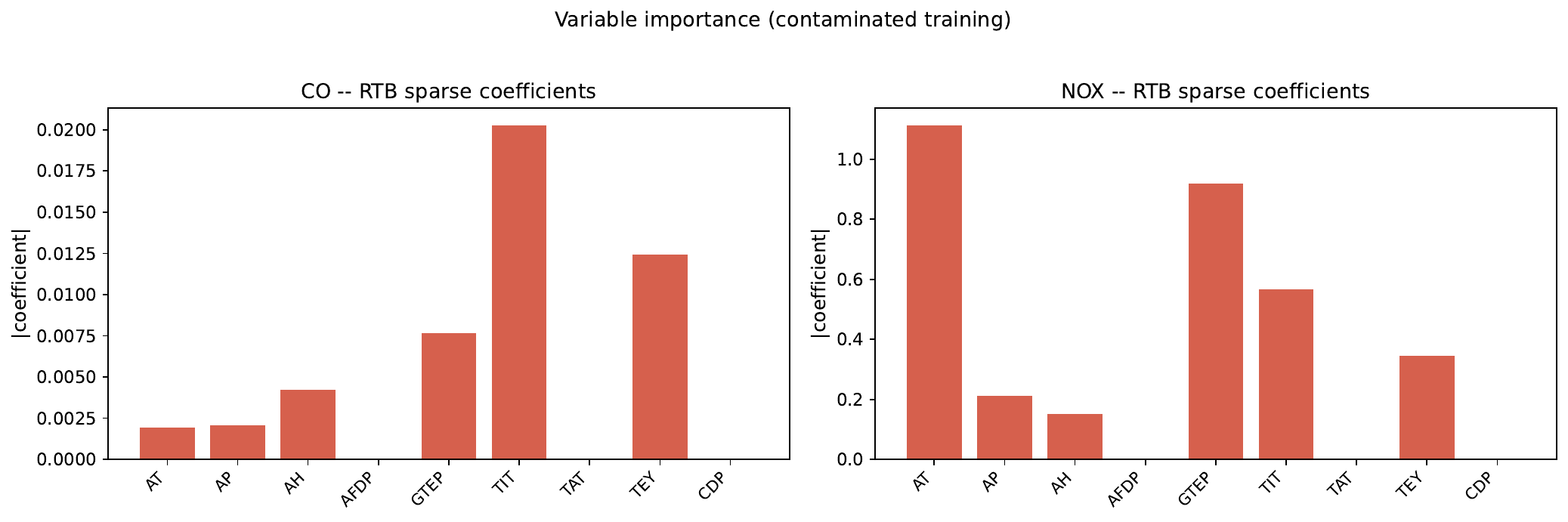}
\caption{RTB sparse absolute coefficient magnitudes for CO (left) and NOx (right) on the contaminated gas turbine data. All 9 predictors are retained; colour intensity reflects coefficient magnitude.}
\label{fig:gas_coefs}
\end{figure}

The sparse RTB model ($\eta_x = 0.7$, $h_x = 5$) eventually does retain all 9 predictor variables on the contaminated data, indicating that all sensor readings carry some information for predicting the emission pair.
However, Figure~\ref{fig:gas_coefs} reveals a strikingly different coefficient profile for each response.
For CO, the dominant predictors are TIT (turbine inlet temperature) and TEY (energy yield), with AFDP, TAT, and CDP contributing near-zero coefficients. For NOx, the dominant predictor is AT (ambient temperature), followed by GTEP (exhaust pressure) and TIT, with a much more distributed coefficient pattern across all variables.
This differential structure illustrates the benefit of multivariate twoblock modelling: the shared latent space captures the common structure, while the per-response coefficients reveal which physical process variables drive each emission type. The fact that the five component RTB model eventually retains all variables, yet still manages to outperform the dense version, indicates that there still is an advantage to extracting components in a sparse manner: for dense RTB, all $\mathbf{X}$ block component weights are non-zero, whereas for sparse twoblock, the component weights are nonzero for just one or a few of the variables. Eventually, the weighted average MSE result shows that there can be a benefit to sparse component extraction, even when the final model ends up using all variables. Moreover, sparse component weights are easier to interpret. 

\section{Conclusions and Outlook}\label{sec:conclusion}

In this article, Robust Twoblock Dimension Reduction (RTB) was introduced. RTB is the first robust variant of two-block simultaneous dimension reduction, available in both dense and sparse forms. It thereby is the first method for simultaneous twoblock dimension reduction that allows to tune the model complexity in each block of variables individually and is robust to outliers. The sparse version is also the first \added[R1.1]{robust} sparse twoblock dimension reduction method that allows to tune a separate number of \replaced[R1.4]{componenents}{latent variables} in each block, as well as a separate sparsity parameter in each block. At this point, it is noted that tuning a separate sparsity parameter was already possible in sparse robust canonical correlation analysis \citep{wilms2016}, yet an intrinsic drawback to canonical correlation analysis is that the same number of \replaced[R1.4]{components}{latent components} needs to apply to both blocks of variables, even in cases where dimensionalities of both blocks are very different. \added[R1.1, R1.D]{Likewise, the non-robust multiblock framework of RGCCA and SGCCA \citep{tenenhaus2014,tenenhaus2017} allows block-specific numbers of components and sparsity, but it targets a symmetric association criterion rather than a regression model and is not resistant to outliers (Section~\ref{sec:related}).} 

By applying iteratively reweighted estimation with independent case weighting for the $\mathbf{X}$ and $\mathbf{Y}$ blocks, RTB achieves resistance to leverage points, vertical outliers, and joint contamination while maintaining estimation efficiency close to classical twoblock on clean data. 

This was illustrated in an extensive simulation study, as well as in a set of two examples. The simulation study demonstrates that RTB consistently outperforms classical twoblock under contamination\added[R1.16]{, in terms of coefficient estimation, prediction and variable selection}, with the benefit increasing with the proportion of outliers, and that these gains extend to the high-dimensional $p > n$ setting.

The work proposed in this article leaves several avenues open to explore in future work. While the method allows to select different sparsities in each block of variables, the focus of this paper was on the robustness properties in a setting of dimension reduction and multivariate regression, as well as on variable selection in the $\mathbf{X}$ block. Variable selection in the block of dependent variables needs to be investigated in future work in a similar way. Applications for that aspect of the method can arise in bioinformatics, for instance in models that link genome data to metabolome data, where it is unknown before the analysis which variables to select in each of these blocks, while outliers can naturally occur in either block. Another area to explore is extension of this method to time series data. While many potential paths for future research can be envisaged, hopefully the results presented in this article can already spur broader adoption of the robust twoblock dimension reduction method.   

\appendix

\section{Code availability}

All methods proposed in this article, are implemented in the \textsf{Python} 3 package \href{https://pypi.org/project/twoblock/}{\texttt{twoblock}}, which has been made publicly available on PyPI and the source code of which can be retrieved on GitHub. The package implements the methods \texttt{twoblock} and the robust counterpart \texttt{rtb}. Its hyperparameters, discussed in this article, can readily be tuned, since the methods are compatible with the widely used \texttt{scikit-learn} API, supporting common attributes, such as \texttt{fit}, \texttt{predict} and \texttt{transform}.

\section{Data availability}

All data are public domain and for convenience, a copy is distributed with the \texttt{twoblock} package in the \texttt{data/} directory (\url{https://github.com/SvenSerneels/twoblock/data}). The original gas turbine CO and NOx emission data set \citep{kaya2019} is publicly available from the UCI Machine Learning Repository at \url{https://archive.ics.uci.edu/dataset/551}. All simulation code and example notebooks used to produce the results in this paper are included in the package repository.

\section{Generative AI use}

In the preparation of this manuscript, Claude Opus 4.6 (Anthropic, Inc.) was used to embellish code and Jupyter notebooks, as well as to transfer some results over into tables and figures in the paper.

\bibliographystyle{apalike}
\bibliography{references}

\begin{thebibliography}{}

\bibitem[Bougeard et~al., 2008]{bougeard2008}
Bougeard, S., Hanafi, M., and Qannari, E.~M. (2008).
\newblock Continuum redundancy--{PLS} regression: A simple continuum approach.
\newblock {\em Computational Statistics \& Data Analysis}, 52(7):3686--3696.

\bibitem[Bougeard et~al., 2011]{bougeard2011}
Bougeard, S., Qannari, E.~M., Lupo, C., and Hanafi, M. (2011).
\newblock From multiblock partial least squares to multiblock redundancy
  analysis. {A} continuum approach.
\newblock {\em Informatica}, 22(1):11--26.

\bibitem[Chessel and Hanafi, 1996]{chessel1996}
Chessel, D. and Hanafi, M. (1996).
\newblock Analyses de la co-inertie de {K} nuages de points.
\newblock {\em Revue de Statistique Appliqu{\'e}e}, 44(2):35--60.

\bibitem[Chun and Kele{\c{s}}, 2010]{chun2010}
Chun, H. and Kele{\c{s}}, S. (2010).
\newblock Sparse partial least squares regression for simultaneous dimension
  reduction and variable selection.
\newblock {\em Journal of the Royal Statistical Society: Series B},
  72(1):3--25.

\bibitem[Cook et~al., 2023]{cook2023}
Cook, R.~D., Forzani, L., and Liu, L. (2023).
\newblock Partial least squares for simultaneous reduction of response and
  predictor vectors in regression.
\newblock {\em Journal of Multivariate Analysis}, 196:105163.

\bibitem[Cook et~al., 2013]{cook2013}
Cook, R.~D., Helland, I.~S., and Su, Z. (2013).
\newblock Envelopes and partial least squares regression.
\newblock {\em Journal of the Royal Statistical Society: Series B (Statistical
  Methodology)}, 75(5):851--877.

\bibitem[de~Jong and Kiers, 1992]{dejong1992}
de~Jong, S. and Kiers, H. A.~L. (1992).
\newblock Principal covariates regression: Part {I}. {T}heory.
\newblock {\em Chemometrics and Intelligent Laboratory Systems}, 14:155--164.

\bibitem[Filzmoser et~al., 2020]{FILZMOSER2020106944}
Filzmoser, P., Höppner, S., Ortner, I., Serneels, S., and Verdonck, T. (2020).
\newblock Cellwise robust m regression.
\newblock {\em Computational Statistics \& Data Analysis}, 147:106944.

\bibitem[Fritz et~al., 2012]{fritz2012comparison}
Fritz, H., Filzmoser, P., and Croux, C. (2012).
\newblock A comparison of algorithms for the multivariate l 1-median.
\newblock {\em Computational Statistics}, 27(3):393--410.

\bibitem[Galharret et~al., 2026]{galharret2026}
Galharret, J.-M., Bougeard, S., Van~Deun, K., and Hanafi, M. (2026).
\newblock Robust canonical factorization of multiblock data.
\newblock In {\em 57es Journ{\'e}es de Statistique de la SFdS},
  Clermont-Ferrand, France.
\newblock HAL: hal-05616577.

\bibitem[Hoffmann et~al., 2015]{hoffmann2015}
Hoffmann, I., Serneels, S., Filzmoser, P., and Croux, C. (2015).
\newblock Sparse partial robust {M} regression.
\newblock {\em Chemometrics and Intelligent Laboratory Systems}, 149:50--59.

\bibitem[Kaya et~al., 2019]{kaya2019}
Kaya, H., T{\"u}fek{\c{c}}i, P., and Uzun, E. (2019).
\newblock Gas turbine {CO} and {NOx} emission data set.
\newblock UCI Machine Learning Repository.
\newblock Dataset \#551, \url{https://archive.ics.uci.edu/dataset/551}.

\bibitem[Maronna et~al., 2019]{maronna2019robust}
Maronna, R.~A., Martin, R.~D., Yohai, V.~J., and Salibi{\'a}n-Barrera, M.
  (2019).
\newblock {\em Robust statistics: theory and methods (with R)}.
\newblock John Wiley \& Sons.

\bibitem[Maronna and Zamar, 2002]{maronna2002}
Maronna, R.~A. and Zamar, R.~H. (2002).
\newblock Robust estimates of location and dispersion for high-dimensional
  datasets.
\newblock {\em Technometrics}, 44(4):307--317.

\bibitem[Osborne et~al., 1984]{osborne1984}
Osborne, B.~G., Fearn, T., Miller, A.~R., and Douglas, S. (1984).
\newblock Application of near infrared reflectance spectroscopy to the
  compositional analysis of biscuits and biscuit doughs.
\newblock {\em Journal of the Science of Food and Agriculture}, 35(1):99--105.

\bibitem[Park et~al., 2021]{park2021}
Park, S., Ceulemans, E., and Van~Deun, K. (2021).
\newblock Sparse common and distinctive covariates regression.
\newblock {\em Journal of Chemometrics}, 35(2):e3270.

\bibitem[Park et~al., 2024]{park2024}
Park, S., Ceulemans, E., and Van~Deun, K. (2024).
\newblock Variable selection for both outcomes and predictors: sparse
  multivariate principal covariates regression.
\newblock {\em Machine Learning}, 113:7319--7370.

\bibitem[Raymaekers and Rousseeuw, 2019]{raymaekers2019generalized}
Raymaekers, J. and Rousseeuw, P. (2019).
\newblock A generalized spatial sign covariance matrix.
\newblock {\em Journal of Multivariate Analysis}, 171:94--111.

\bibitem[Rousseeuw, 1984]{rousseeuw1984}
Rousseeuw, P.~J. (1984).
\newblock Least median of squares regression.
\newblock {\em Journal of the American Statistical Association},
  79(388):871--880.

\bibitem[Rousseeuw, 1985]{rousseeuw1985}
Rousseeuw, P.~J. (1985).
\newblock Multivariate estimation with high breakdown point.
\newblock In Grossmann, W., Pflug, G., Vincze, I., and Wertz, W., editors, {\em
  Mathematical Statistics and Applications, Vol.\ B}, pages 283--297. Reidel
  Publishing Company, Dordrecht.

\bibitem[Serneels, 2025]{serneels2025}
Serneels, S. (2025).
\newblock Sparse twoblock dimension reduction: A versatile alternative to
  sparse {PLS2} and {CCA}.
\newblock {\em Journal of Chemometrics}, 39:e70051.

\bibitem[Serneels et~al., 2005]{serneels2005}
Serneels, S., Croux, C., Filzmoser, P., and Van~Espen, P.~J. (2005).
\newblock Partial robust {M}-regression.
\newblock {\em Chemometrics and Intelligent Laboratory Systems},
  79(1--2):55--64.

\bibitem[Serneels et~al., 2024]{serneels2024}
Serneels, S., Insolia, L., and Verdonck, T. (2024).
\newblock Elegant robustification of sparse partial least squares by
  robustness-inducing transformations.
\newblock {\em Statistics}, 58(1):44--64.

\bibitem[Smilde et~al., 2017]{smilde2017}
Smilde, A.~K., M{\aa}ge, I., N{\ae}s, T., Hankemeier, T., Lips, M.~A., Kiers,
  H. A.~L., Acar, E., and Bro, R. (2017).
\newblock Common and distinct components in data fusion.
\newblock {\em Journal of Chemometrics}, 31(7):e2900.

\bibitem[Tenenhaus et~al., 2014]{tenenhaus2014}
Tenenhaus, A., Philippe, C., Guillemot, V., L{\^e}~Cao, K.-A., Grill, J., and
  Frouin, V. (2014).
\newblock Variable selection for generalized canonical correlation analysis.
\newblock {\em Biostatistics}, 15(3):569--583.

\bibitem[Tenenhaus and Tenenhaus, 2011]{tenenhaus2011}
Tenenhaus, A. and Tenenhaus, M. (2011).
\newblock Regularized generalized canonical correlation analysis.
\newblock {\em Psychometrika}, 76(2):257--284.

\bibitem[Tenenhaus et~al., 2017]{tenenhaus2017}
Tenenhaus, M., Tenenhaus, A., and Groenen, P. J.~F. (2017).
\newblock Regularized generalized canonical correlation analysis: A framework
  for sequential multiblock component methods.
\newblock {\em Psychometrika}, 82(3):737--777.

\bibitem[Tucker, 1958]{tucker1958}
Tucker, L.~R. (1958).
\newblock An inter-battery method of factor analysis.
\newblock {\em Psychometrika}, 23(2):111--136.

\bibitem[van~den Berg et~al., 2006]{vandenberg2006}
van~den Berg, R.~A., Hoefsloot, H. C.~J., Westerhuis, J.~A., Smilde, A.~K., and
  van~der Werf, M.~J. (2006).
\newblock Centering, scaling, and transformations: improving the biological
  information content of metabolomics data.
\newblock {\em BMC Genomics}, 7:142.

\bibitem[van Rijsbergen, 1979]{vanrijsbergen1979}
van Rijsbergen, C.~J. (1979).
\newblock {\em Information Retrieval}.
\newblock Butterworths, London, 2nd edition.

\bibitem[Wilms and Croux, 2016]{wilms2016}
Wilms, I. and Croux, C. (2016).
\newblock Robust sparse canonical correlation analysis.
\newblock {\em BMC Systems Biology}, 10:1--13.

\bibitem[Witten et~al., 2009]{witten2009}
Witten, D.~M., Tibshirani, R., and Hastie, T. (2009).
\newblock A penalized matrix decomposition, with applications to sparse
  principal components and canonical correlation analysis.
\newblock {\em Biostatistics}, 10(3):515--534.

\bibitem[Wold, 1966]{wold1966}
Wold, H. (1966).
\newblock Nonlinear estimation by iterative least squares procedures.
\newblock In David, F., editor, {\em Papers in Statistics: Festschrift for J.
  Neyman}, pages 411--444. Wiley.

\end{thebibliography}

\end{document}